\documentclass[twocolumn]{aa}

\usepackage{txfonts}
\usepackage{graphicx}

\begin{document}

\title{3D WKB solution for fast magnetoacoustic wave behaviour around an X-line}

\author{J.~A.~{McLaughlin},  G.~J.~J.~{Botha}, S.~{R\'{e}gnier} \and D.~L.~{Spoors}}

\offprints{J.~A.~McLaughlin, \email{james.a.mclaughlin@northumbria.ac.uk}}

\institute{Department of Mathematics and Information Sciences, Northumbria University, Newcastle upon Tyne, NE1 8ST, UK}

\date{Received; Accepted}

\authorrunning{McLaughlin {{et al.}}}
\titlerunning{3D WKB solution for fast magnetoacoustic wave behaviour around an X-Line}

%----------------------------------------------------------------

\abstract
{We study the propagation of a fast magnetoacoustic wave in a 3D magnetic field created from two magnetic dipoles. The magnetic topology contains an X-line.}
{We  aim to  contribute to the overall understanding of MHD wave propagation within inhomogeneous media, specifically around X-lines.}
{We investigate the linearised, 3D MHD equations under the assumptions of ideal and cold plasma. We utilise the WKB approximation and  Charpit's method  during our investigation.}
{It is found that the behaviour of the fast magnetoacoustic wave is entirely dictated by the local, inhomogeneous, equilibrium Alfv\'en speed profile. All parts of the wave experience refraction during propagation, where the  magnitude of the refraction effect  depends on the location of an individual wave element within the inhomogeneous magnetic field. The X-line, along which the  Alfv\'en speed is identically zero, acts as a focus for the refraction effect. There are two main types of wave behaviour:  part of the wave is either trapped by the X-line  or escapes the system, and there exists a  critical starting region around the X-line that divides these two types of behaviour. For the set-up investigated, it is found that  $15.5\%$ of the fast wave energy is trapped by the X-line.}
{We conclude that linear, $\beta=0$ fast magnetoacoustic waves can accumulate along X-lines and thus these will be specific  locations of fast wave energy deposition and thus preferential heating. The work here highlights the importance of understanding the magnetic topology of a system. We also demonstrate how the 3D WKB technique  described in this paper can  be applied to other magnetic configurations.}

\keywords{Magnetohydrodynamics (MHD); Magnetic fields; Waves; Sun: corona; Sun: magnetic fields; Sun: oscillations}

\maketitle

%-------------------------------------------------

\section{Introduction}\label{introduction_chapter}

It is now clear that magnetohydrodynamic (MHD) wave motions (e.g.  Roberts \cite{Bernie};  Nakariakov \& Verwichte \cite{NV2005}; De Moortel \cite{DeMoortel2005}) are  ubiquitous throughout the solar atmosphere (Tomczyk et al. \cite{Tomczyk}). Several different types of MHD wave motions have been observed by various solar instruments: longitudinal propagating disturbances  have been seen in {{SOHO}} data (e.g. Berghmans \& Clette \cite{Berghmans1999}; Kliem {{et al.}} \cite{Kliem}; Wang {{et al.}} \cite{Wang2002}) and {{TRACE}} data (De Moortel {{et al.}} \cite{DeMoortel2000}) and these have been interpreted as slow magnetoacoustic  waves. Transverse waves have been observed in the corona and chromosphere with {{TRACE}} (Aschwanden {{et al.}} \cite{Aschwandenetal1999}, \cite{Aschwandenetal2002}; Nakariakov {{et al.}} \cite{Nakariakov1999}; Wang \& Solanki \cite{Wang2004}), {{Hinode}} (Okamoto {{et al.}} \cite{Okamoto}; De Pontieu {{et al.}} \cite{Bart2007}; Ofman \& Wang \cite{OW2008}), SDO data  (e.g. McIntosh et al. \cite{McIntosh2011}; Morton et al.  \cite{Morton2012}, \cite{Morton2015}; Morton \& McLaughlin \cite{MortonMcLaughlin2013}, \cite{MortonMcLaughlin2014}; Thurgood et al. \cite{Thurgood2014}) and these have been interpreted as fast magnetoacoustic waves, specifically kink waves. These transverse motions have also been interpreted as Alfv\'enic waves, although  this interpretation is subject to  discussion, e.g. see Erd{\'e}lyi \& Fedun (\cite{RF2007}), Van Doorsselaere {{et al.}} (\cite{Tom2008}) and Goossens et al. (\cite{Goossens2009}). Non-thermal line  broadening due to torsional Alfv\'en waves has been reported by  {Erd{\'e}lyi} {{et al.}} (\cite{E1998}), Harrison {{et al.}} (\cite{Harrison2002}),  O'Shea {{et al.}} (\cite{Oshea}) and   Jess {{et al.}} (\cite{Jess2009}).

It is also clear that the  coronal magnetic field  plays a fundamental role in the  propagation and properties of MHD waves, and to begin to understand this  inhomogeneous magnetised  environment it is useful to look at the topology (structure) of the magnetic field itself.  Potential-field extrapolations of the coronal magnetic field can be made from photospheric magnetograms (e.g. see R{\'e}gnier \cite{Stephane2013}) and such extrapolations show the existence of important features of the topology: {\emph{null points}} - specific points  where the magnetic field is zero, {\emph{separatrices}} - topological features that separate regions of different magnetic flux connectivity, and {\emph{X-lines}} or {\emph{null lines}} - extended locations  where the magnetic field, and thus the Alfv\'en speed, is zero. Investigations of the coronal magnetic field using such potential field calculations can be found in, e.g.,  Brown \& Priest (\cite{BrownPriest2001}),  Beveridge et al. (\cite{Beveridge2002}), R{\'e}gnier et al. (\cite{Stephane2008}) and in a comprehensive review  by Longcope (\cite{L2005}).

These two areas of scientific study, namely  ubiquitous MHD waves  and magnetic topology, will naturally encounter each other in the solar atmosphere, e.g.  MHD waves will propagate into the neighbourhood of coronal null points, X-lines and separatrices. Thus, the study of MHD waves within inhomogeneous magnetic media is  itself a fundamental physical process. Previous works, detailed below, have focused on MHD wave behaviour in the neighbourhood of null points and separatrices (see review by McLaughlin et al. \cite{McLaughlinREVIEW}). However, less attention has been given to the transient behaviour of MHD waves in the vicinity of X-lines in the solar atmosphere. The motivation for this paper is to address this, i.e. this paper aims to investigate the behaviour of fast MHD waves around an X-line in order to contribute to the overall understanding of MHD wave propagation within inhomogeneous media. {{Note that an X-line is a degenerate structure and its existence requires a special symmetry of the magnetic field. Thus, given the inherent lack of symmetry in solar magnetic observations, their existence in the solar atmosphere is unlikely. However, }} X-lines are well studied in other areas, such as in the Earth's magnetosphere,  e.g. Runov et al. (\cite{Runov2003}) and Phan et al. (\cite{Phan2006}).

The propagation of fast magnetoacoustic waves in an inhomogeneous coronal plasma has been investigated by Nakariakov \& Roberts (\cite{Nakariakov1995}), who showed that the waves are refracted into regions of low Alfv\'en speed (see also Thurgood \& McLaughlin \cite{Thurgood}). In the case of X-lines, the Alfv\'en speed actually drops to zero.

MHD waves in the neighbourhood of a single 2D X-point have been investigated by various authors. Bulanov \& Syrovatskii (\cite{Bulanov1980}) provided a detailed discussion of the propagation of  fast and Alfv\'en waves using cylindrical symmetry. Craig \& Watson (\cite{CraigWatson1992}) mainly considered the radial propagation of the $m=0$ mode (where $m$ is the azimuthal wavenumber) using a mixture of analytical and numerical solutions. They showed that the propagation of the $m=0$ wave towards the null point generates an exponentially large increase in the current density. Craig \& McClymont (\cite{CraigMcClymont1991}, \cite{CraigMcClymont1993}), Hassam (\cite{Hassam1992}) and Ofman et al. (\cite{OMS1993}) investigated the normal mode solutions for both $m=0$ and $m\ne 0$ modes with resistivity included. They emphasise that the current builds as the inverse square of the radial distance from the X-point. All these investigations were carried out using cylindrical models in which the generated waves encircled the X-point and so the cylindrical symmetry meant that the disturbances can only propagate either towards or away from the X-point. The behaviour of MHD waves around two-dimensional X-points in a Cartesian geometry has been investigated by McLaughlin \& Hood (\cite{MH2004}, \cite{MH2005}, \cite{MH2006b}),  McLaughlin et al. (\cite{MDHB2009}) and more recently by {Ku{\'z}ma} et al. ({\cite{Kuzma2015}). Of note is also McLaughlin \& Hood (\cite{MH2006a}) who investigated fast MHD wave propagation in the neighbourhood of two dipoles. These authors  solved the linearised, $\beta=0$  MHD equations and found that   the propagation of the  linear fast wave is dictated by the  Alfv\'en speed profile and that close to the X-point, the  wave is attracted to the X-point by a refraction effect. It was also found that  in this magnetic configuration a proportion of the wave can  escape the refraction effect and that the split  occurs  near the regions of very high  Alfv\'en speed. However, this study was limited to 2D. The current paper extends this work to 3D.

MHD waves in the vicinity of a 3D null point (e.g. Parnell et al. \cite{Parnell1996}; Priest \& Forbes \cite{PF2000}) have also been investigated. Galsgaard et al. (\cite{Galsgaard2003}) performed numerical experiments on the effect of twisting the spine of a 3D null point, and described the resultant wave propagation towards the null. They found that when the fieldlines around the spine are perturbed in a rotationally symmetric manner, a twist wave (essentially an Alfv\'en wave) propagates towards the null along the fieldlines.  Whilst this Alfv\'en wave spreads out as the null is approached, a fast-mode wave focuses on the null point and wraps around it. In addition,  Pontin \& Galsgaard (\cite{PG2007}) and Pontin et al. (\cite{PBG2007}) performed numerical simulations in which the spine and fan of a 3D null point are subject to rotational and shear perturbations. They found that  rotations of the fan plane lead to current sheets in the location of the spine and rotations about the spine lead to current sheets  in the fan.

The WKB approximation is an asymptotic approximation technique which can be used when a system contains a large parameter (see {{e.g.}} Bender \& Orszag \cite{Bender1978}). Hence, the WKB method can be used in a system where a wave propagates through a background medium which varies on some spatial scale which is much longer than the wavelength of the wave. There are several examples of authors utilising the WKB approximation to compare with numerical results, e.g. Khomenko \& Collados (\cite{Khomenko2006}) and Afanasyev \& Uralov (\cite{Afanasyev2011}, \cite{Afanasyev2012}). Galsgaard et al. (\cite{Galsgaard2003}) compared their numerical results with a WKB approximation and found that, for the $\beta=0$ fast wave, the wavefront wraps around the null point as it contracts towards it. They perform their WKB approximation in cylindrical polar coordinates and thus their resultant equations are two-dimensional, since a simple 3D null point is essentially 2D in cylindrical coordinates.  In contrast, this paper will solve the WKB equations for three Cartesian components, and thus we can solve for more general disturbances and more general boundary conditions. McLaughlin et al. (\cite{MFH2008}) utilised the WKB approximation to investigate MHD wave behaviour in the neighbourhood of a fully 3D null point. The authors utilised the WKB approximation to determine the transient properties of the fast and Alfv\'en modes in a linear, $\beta=0$ plasma regime.  From these works, it has been  demonstrated that  the WKB approximation can provide a vital link between analytical and numerical work, and often provides the critical insight  into understanding the physical results. This paper demonstrates the methodology of  how to apply the WKB approximation to a general 3D magnetic field configuration. We believe that with the vast amount of 3D modelling currently being undertaken, applying this WKB technique in 3D will be very useful and beneficial to the MHD modelling community.

This paper describes an investigation into the behaviour of fast MHD waves around an X-line  using the WKB approximation. The paper has the following outline: In $\S\ref{SEC:1}$, the basic equations, linearisation and assumptions are described, including details of our equilibrium magnetic field. $\S\ref{WKBAPPROXIMATION}$ details the WKB technique utilised in this paper as well as its application to the fast wave. The results are given in $\S\ref{sec:results}$ and the conclusions  and discussion are presented in $\S\ref{conclusion}$. There are multiple appendices (\ref{appendixB}, \ref{appendixA}, \ref{appendixC}) which complement the work in the core text.

%%%%%%%%%%%%%%%%%%%%%%%%%%%%%%%%%%%%%%%%%%%%%%%%%%%%%%%%%%%%%%%%%%%%%%%%%%%%%%%%%%%%%%

\begin{figure*}[t]
\begin{center}
\includegraphics[width=5.35in]{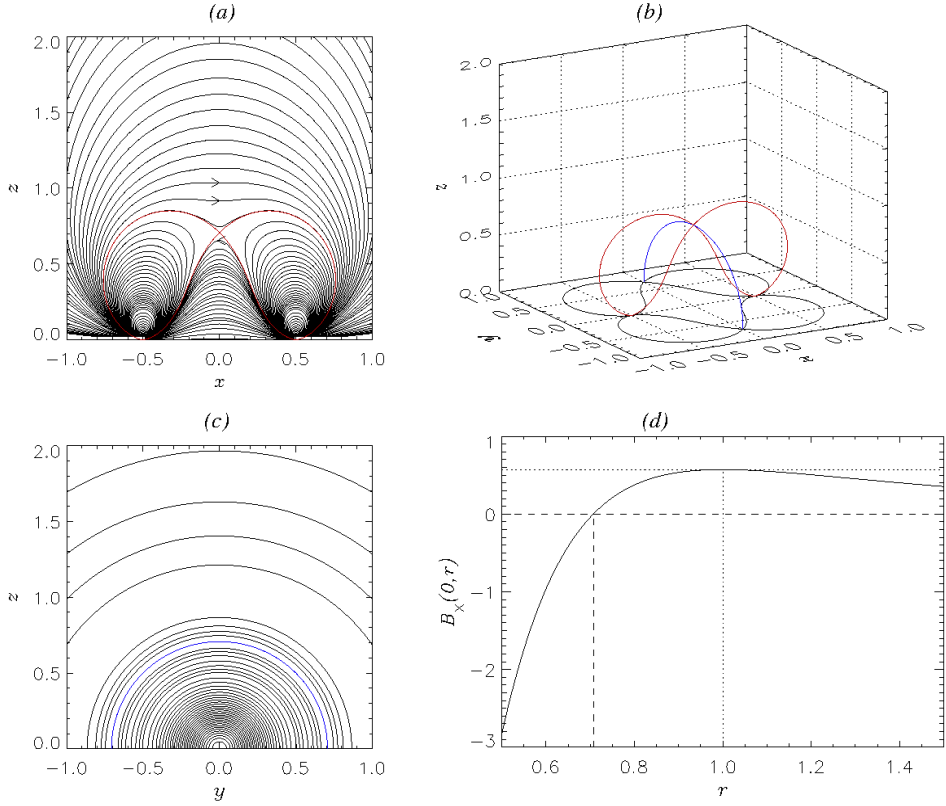}
\caption{$(a)$ Equilibrium magnetic field  in the $y=0$, $xz-$plane. Dipoles are located at $x=\pm 0.5$. X-point is located at $x=0$ and $z=\sqrt{2}a=\sqrt{0.5}=0.707$.  Red lines indicate the separatrices in this plane.  $(b)$ 3D visualisation of the  equilibrium magnetic field denoting the red separatrices along $y=0$ from $(a)$ and perpendicular to this the blue X-line along $x=0$ from $(c)$. Equilibrium magnetic field is rotationally symmetric about the $y=0$ axis and thus black curves denote the separatrices in the $xy-$plane at $z=0$. $(c)$  Equilibrium magnetic {{field}} shown in the $x=0$, $yz-$plane. Magnetic field is only in the $x-$direction, hence no arrows. Blue line denotes the X-line of the form $y^2+z^2=2a^2$.  $(d)$  Plot of $B_x(0,r)$ where $r^2=y^2+z^2$.  $B_x(0,r)$ changes sign at $r = \sqrt{2}a= \sqrt{0.5}=0.707$, i.e. at location of the X-line. Maximum of $d B_x(0,r) / dr$ occurs at $r=1$, where  $B_x(0,r=1)= (4/5)^{5/2} = 0.5724$.}
\label{Figure1}
\end{center}
\end{figure*}

\begin{figure*}[t]
\begin{center}
\includegraphics[width=5.35in]{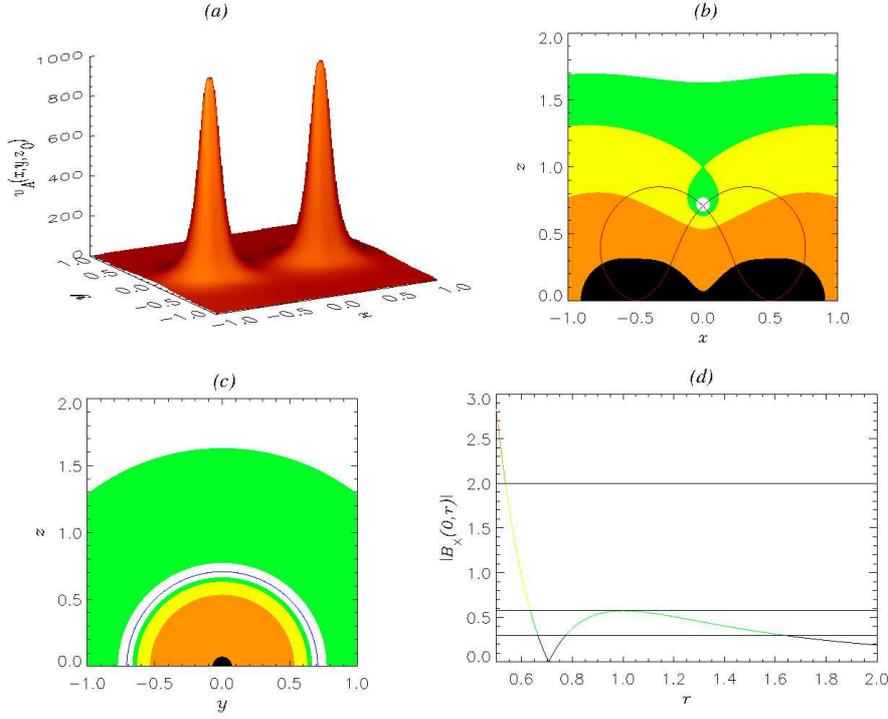}
\caption{$(a)$  Shaded surface of  $v_A(x,y,z_0) = | {\bf{B}}_0(x,y,z_0)|$ in the $xy-$plane at $z_0=0.1$, with local maxima at  $(x,y,z)=(\pm a,0,0)$, i.e. the dipoles' location. $(b)$ Colour contour of $v_A(x,0,z) = | {\bf{B}}_0(x,0,z)|$ in the $y=0$, $xz-$plane. Contour is  colour coded:  $0 \le v_A \le 0.3$ (white);  $0.3 \le v_A \le 0.5724$ (green);  $0.5724 \le v_A \le 2$ (yellow);  $2 \le v_A \le 30$ (orange);   $v_A \ge 30$ (black). Red lines indicate the separatrices in this plane.  $(c)$  Colour contour of $v_A(0,y,z) = | {\bf{B}}_0(0,y,z)|$ in the $x=0$, $yz-$plane. Blue line indicates the X-line in this plane. Contour is  colour coded in the same way as $(b)$. $(d)$ Plot of  $|B_x(0,r)|=v_A(0,y,z)$ where $r^2+y^2+z^2$ and axis displays $0.5 \le r \le 2$. Colour coding corresponds to that of  $(b)$ and $(c)$, except now  black represents  $|B_x(0,r)| \le 0.3$.}
\label{Figure2}
\end{center}
\end{figure*}

%%%%%%%%%%%%%%%%%%%%%%%%%%%%%%%%%%%%%%%%%%%%%%%%%%%%%%%%%%%%%%%%%%%%%%%%%%%%%%%%%%%%%%

\section{Governing equations}\label{SEC:1}

\subsection{Basic equations}\label{SEC:1.1}

The resistive, adiabatic MHD equations for a plasma in the solar corona are used:
\begin{eqnarray}
\rho{\partial {\bf v} \over \partial t} + \rho \left ({\bf v}\cdot \nabla \right){\bf v} &=&  - \nabla p + {\bf j} \times {\bf B} + \rho {\bf{g}}  \nonumber \;,\\
{\partial {\bf B} \over \partial t} &=& \nabla \times \left ({\bf v} \times {\bf B} \right) + \eta \nabla^{2} {\bf B}  \;,  \nonumber\\
{\partial {\rho} \over \partial t} + \nabla \cdot \left (\rho{\bf v} \right) &=& 0\;,\nonumber\\
{\partial p \over \partial t} + {\bf v} \cdot \nabla p &=& - \gamma p \nabla \cdot {\bf v}\;,\nonumber\\
{{\mu}_0} \:{\bf{j}} &=& \nabla \times {\bf{B}}\; ,\label{4}
\end{eqnarray}
where ${\bf v}$  is the plasma velocity, ${\rho}$  is the mass density, ${p}$  is the plasma pressure, ${\bf B}$  is the magnetic induction (usually called the magnetic field), ${\bf{j}}$ is the electric current density, ${\bf{g}}$ is gravitational acceleration, ${\gamma}$  is the ratio of {{specific}} heats, ${\eta}$  is the magnetic diffusivity and ${{\mu}_0}$ is the magnetic permeability in a vacuum.

\subsection{Linearised equations and non-dimensionalisation}

In this paper, the linearised MHD equations are used to study the nature of wave propagation. Using subscripts of $0$ for equilibrium quantities and $1$ for perturbed quantities, equations (\ref{4}) become:
\begin{eqnarray}
\rho_{0}{\partial {\bf v}_{1} \over \partial t} &=&  - \nabla p_1 + {\bf {j}}_{0}  \times {{\bf B}}_1 + {{\bf {j}}}_1 \times {\bf B}_{0}  + \rho_1 {\bf{g}}\;, \label{5}\\
{\partial {{\bf B}}_1 \over \partial t} &=& \nabla \times \left ({\bf v}_{1} \times {\bf B}_{0} \right) + \eta \nabla^{2} {{\bf B}}_1   \;,    \label{6}\\
{\partial {\rho_{1}} \over \partial t} &+& \nabla \cdot \left (\rho_{0}{\bf v}_{1} \right) = 0     \;,    \label{7}\\
{\partial p_{1} \over \partial t} &+& {{\bf v}_{1}} \cdot  \nabla p_{0} = - \gamma p_{0} \nabla \cdot {\bf v}_{1}   \;,        \label{7b}\\
{{\mu}_0} \:{\bf{j}}_1 &=& \nabla \times {\bf{B}}_1\; , \label{8}
\end{eqnarray}
{{where we note that ${{\bf {v}}_{0}}={\bf{0}}$}}. We now consider several simplifications to our system. {{We will be  considering a potential equilibrium magnetic field and so $\nabla \times  {\bf B}_{0}={\bf {j}}_{0}={\bf{0}}$. In addition, we ignore the effect of gravity on the system, i.e. we set $ {{g}} = 0$. We also  consider an ideal system and so the  magnetic diffusivity, $\eta$, is set to zero.}}

{{

Furthermore, we consider a cold plasma, i.e. $\beta=0$ plasma approximation, since in the solar corona $\beta \ll 1$. Under this assumption, $p_0=0$ and $p_1=p_1(x,y,z)$ from  equation (\ref{7b}). We will also assume the equilibrium gas density, $\rho_0$, is uniform. Note that a spatial variation in $\rho _0$ can cause phase mixing (e.g. Heyvaerts \& Priest \cite{Heyvaerts1983}; De Moortel {{et al.}} \cite{DeMoortel1999}; McLaughlin {{et al.}} \cite{McLaughlin2011}). There are no assumptions on $\rho_1=\rho_1(x,y,z,t)$ but we will not discuss equation (\ref{7}) further as it can be solved once we know $\mathbf{v} _1$. In fact, under the assumptions of $\beta=0$, linearisation and no gravity, ${\rho_{1}}$ has no influence on the momentum equation  and so the plasma is effectively arbitrarily compressible ({Craig} \& {Watson} \cite{CraigWatson1992}).

}}

We now non-dimensionalise the above equations as follows: let ${\mathbf{\mathrm{{\bf{v}}}}}_1 = \bar{\rm{v}} {\mathbf{v}}_1^*$, ${\mathbf{B}}_0 = B {\mathbf{B}}_0^*$, ${\mathbf{B}}_1 = B{\mathbf{B}}_1^*$, $x = L x^*$,    $y = L y^*$, $z=Lz^*$, $\nabla = \nabla^* / L$ and $t=\bar{t}\:t^*$, where we let $*$ denote a dimensionless quantity and $\bar{\rm{v}}$, $B$,       $L$, and $\bar{t}$ are constants with the dimensions of the variable that they are scaling. In addition, $\rho_0$ and $p_0$ are constants as these equilibrium quantities are uniform, {{i.e.}} $\rho_0^*=p_0^*=1$.  We then set $ {B} / {\sqrt{{{\mu}_0} \rho _0 } } =\bar{\rm{v}}$ and $\bar{\rm{v}} =  {L} / {\bar{t}}$, which sets $\bar{\rm{v}}$ as a constant equilibrium Alfv\'{e}n speed. Under these scalings $t^*=1$, for example, refers to $t=\bar{t}=  {L} / {\bar{\rm{v}}}$, {{i.e.}} the equilibrium Alfv\'en time taken to travel a distance $L$.  For the rest of this paper, we drop the star indices; the fact that they are now non-dimensionalised is understood. Thus, our $\beta=0$, ideal, linearised, non-dimensionalised equations are given by:
\begin{eqnarray}
{\partial {\bf v}_{1} \over \partial t} = \left( \nabla \times {{\bf {B}}}_1\right) \times {\bf B}_{0} \; \quad {\rm{and}} \quad {\partial {{\bf B}}_1 \over \partial t} = \nabla \times \left ({\bf v}_{1} \times {\bf B}_{0} \right)   \;.    \label{linearised_nondimensionalised}
\end{eqnarray}
{{ Note that once $\mathbf{v} _1$ is known, $\rho_1$ can be calculated from  equation (\ref{7}). }}

Equations (\ref{linearised_nondimensionalised}) can be combined to form a single equation:
\begin{eqnarray}
\frac{\partial^2 } {\partial t^2}{\mathbf{v}}_1 =  \left\{ \nabla \times \left[  \nabla \times \left( {\mathbf{v}}_1 \times {\mathbf{B}}_0 \right) \right] \right\} \times {\mathbf{B}}_0   \;. \label{10}
\end{eqnarray}
{{This}} equation is valid for any 3D potential equilibrium magnetic field,  ${\mathbf{B}}_0$. Thus, we now detail our choice of  ${\mathbf{B}}_0$.

%%%%%%%%%%%%%%%%%%%%%%%%%%%%%%%%%%%%%%%%%%%%%%%%%%%%%%%%%%%%%%%%%%

\subsection{Magnetic equilibrium}\label{mageqn}

We choose a magnetic field created by two magnetic dipoles located at $(x,y,z)=  (\pm a, 0, 0)$. The mathematical form of our dipolar magnetic field comes from the vector potential, ${\bf{A}}$, produced by a magnetic dipole moment, ${\bf{m}}$, where  ${\bf{B}}_0 = \nabla \times {\bf{A}}$ and:
\begin{eqnarray*}
{\bf{A}}({\bf{x}})=\frac{{{\mu}_0}}{4\pi}\frac{{\bf{m}}\times{\bf{x}}}{|{\bf{x}}|^3}
\end{eqnarray*}
where ${{{\mu}_0}}$ is the permeability of free space and ${\bf{x}}=(x,y,z)$. See Shadowitz (\cite{Shadowitz1975}) for further details. Thus, the magnetic field takes the form  ${\bf{B}}_0 = \left( B_x, B_y, B_z \right) {B} / L^3$, where:
\begin{eqnarray}
B_x &=&  \frac{ - 2\left( x+a\right)^2 +y^2 + z^2 }{ \left[\left( x+a\right)^2 +y^2+z^2\right]^{5/2}} +   \frac{ - 2\left( x-a\right)^2 + y^2+z^2 }{ \left[\left( x-a\right)^2 +y^2+z^2\right]^{5/2}}\; ,\nonumber\\
B_y &=& -\frac { 3\left( x+a\right)y}{ \left[\left( x+a\right)^2 +y^2 +z^2\right]^{5/2}} -  \frac { 3\left( x-a\right)y}{ \left[\left( x-a\right)^2 +y^2+z^2\right]^{5/2}}\; ,\nonumber\\
B_z &=& -\frac { 3\left( x+a\right)z}{ \left[\left( x+a\right)^2 +y^2 +z^2\right]^{5/2}} -  \frac { 3\left( x-a\right)z}{ \left[\left( x-a\right)^2 +y^2+z^2\right]^{5/2}}\; , \label{two_3D_dipoles}
\end{eqnarray}
where $B$ is a characteristic field strength, $L$ is the length scale for magnetic field variations and the loci of the dipoles are located at $\pm a$ ($2a$ is the separation of the dipoles). We choose $a=0.5L$ in our investigation and so $a=0.5$ under our non-dimensionalisation. The magnetic field can be seen in Figure \ref{Figure1}. The equilibrium magnetic field comprises of separatrix surfaces, i.e. the magnetic skeleton, that divide the magnetic region into four topologically distinct regions. Figure \ref{Figure1}a shows the  equilibrium magnetic field in the $xz-$plane at $y=0$ where the red lines  indicate the separatrices in this plane. Note that the  equilibrium magnetic field is rotationally symmetric about the axis $y=0$, and so the magnetic field geometry in the $xy-$plane along $z=0$ is identical to that of Figure \ref{Figure1}a for $z \rightarrow y$  for $y \ge 0$ and  $z \rightarrow -y$ for $y \le 0$, respectively. This symmetry between $y$ and $z$ can also be understood from the form of the equations for $B_y$ and $B_z$ themselves, which are identical under the mapping  $\left( y,z \right) \rightarrow \left( z,y \right)$.

The (red) separatrices cross at an X-point, located at $(x,y,z)=(0,0,\sqrt{2}a)$ and at that location $B_x(0,0,\sqrt{2}a)=B_y(0,0,\sqrt{2}a)=B_z(0,0,\sqrt{2}a)=0$. This X-point, in the $xz-$plane at $y=0$, forms an {{X-line}} of the form $y^2+z^2=2a^2$ in the $yz-$plane at $x=0$. This can be seen in Figure \ref{Figure1}b where the blue line denotes the X-line. The X-line is central to the investigation in this paper. Note that the magnetic field is identically zero along the whole of the X-line. There is no guide-field along the X-line. Note that the X-point in the $y=0$, $xz-$plane is just a cut across the X-line, and so  $y^2+z^2 \Rightarrow  z^2 =2a^2 \Rightarrow z=\sqrt{2}a$ at $x=y=0$.

Along $x=0$, equations (\ref{two_3D_dipoles}) simplify greatly such that:
\begin{eqnarray}
B_x(0,y,z) &=&  \frac{ 2 \left(- 2a^2 +y^2 + z^2 \right)}{ \left(a^2 +y^2+z^2\right)^{5/2}} \;,\nonumber\\
B_y(0,y,z) &=& B_z(0,y,z)=0\; , \label{X-line_equation_beta_ray_bill}
\end{eqnarray}
where along  $y^2+z^2=2a^2$,  $B_x=B_y=B_z=0$. Figure \ref{Figure1}c  shows the  equilibrium magnetic field in the $yz-$plane at $x=0$ where the blue curve denotes the X-line. Note that when $x=0$,  the magnetic field is in the $x-$direction only, is orientated  perpendicular to the $yz-$plane and, crucially,  changes direction across the X-line. The X-line manifests as a circle, $y^2+z^2=2a^2$,  which is in agreement with our observation of rotational symmetry about the $y=0$ axis.

Letting  $r^2=y^2+z^2$ simplifies equation (\ref{X-line_equation_beta_ray_bill}) further:
\begin{eqnarray}
B_x(0,r) &=&  \frac{ 2 \left(- 2a^2 + r^2 \right)}{ \left(a^2 +r^2\right)^{5/2}} \;,\nonumber\\
B_y(0,r) &=& B_z(0,r)=0\; . \label{X-line_equation_2}
\end{eqnarray}
Thus, in Figure \ref{Figure1}c and inside the blue circle, $B_x<0$ and so the magnetic field is in the negative $x-$direction (into the page), whereas outside the circle  $B_x>0$ and the magnetic field is in the $x-$direction (towards the reader).

Figure \ref{Figure1}d shows a plot of $B_x(0,r)$. We see that $B_x(0,r)$ changes sign as it passes through  $r = \sqrt{2}a= \sqrt{0.5}$ as expected, i.e. this is the location of the X-line. {{There}} is a maximum at $r=1$, i.e.  $\max \left[ B_x(0,r=1) \right]= (4/5)^{5/2} = 0.5724$.

Note that as $x$, $y$ or $z$ get very large, the field strength becomes small; this is a more physically-realistic topology than those previously investigated in  McLaughlin {{et al.}}  (\cite{MFH2008}). Note that although the magnetic field is inhomogeneous, it is still both potential, $\nabla \times {\bf{B}}_0 ={\bf{0}}$, and solenoidal, $\nabla \cdot {\bf{B}}_0 =0$.

%%%%%%%%%%%%%%%%%%%%%%%%%%%%%%%%%%%%%%%%%%%%%%%%%%%%%%%%%%%%%%%%%%%%%%%%%%%%%%%%%%%%%%%%%%%%%%%%%%%%%%

\subsection{Alfv\'en speed profile}

Previous work (see McLaughlin et al. \cite{McLaughlinREVIEW}) has highlighted that the equilibrium Alfv\'en speed profile,  $v_A(x,y,z)=|{\bf{B}}_0(x,y,z)|$, plays a key role in dictating the propagation of the fast wave. Figure \ref{Figure2}a shows a shaded surface of  $v_A(x,y,0) = | {\bf{B}}_0(x,y,0)|$ in the $xy-$plane {{at $z=0.1$}}. The shaded surface clearly shows that  the equilibrium Alfv\'en speed profile changes substantially across the magnetic region and reaches maxima at  $(x,y,z)=(\pm a,0,0)$, i.e. the location of the dipoles. Figure \ref{Figure2}b shows a colour contour of $v_A(x,0,z) = | {\bf{B}}_0(x,0,z)|$ along in the $xz-$plane at $y=0$. As in Figure \ref{Figure1}a, the red lines indicate the separatrices in this plane. The contour is  colour coded: white represents values  $0 \le v_A \le 0.3$; green represents  $0.3 \le v_A \le 0.5724$ (where $\max \left[ B_x(0,r) \right] = 0.5724$ at $r=1$); yellow represents  $0.5725 \le v_A \le 2$; orange represents  $2 \le v_A \le 30$; and black represents  $v_A \ge 30$. Thus, we can see that around the X-point, located at $x=0$ and $z=\sqrt{2}a=\sqrt{0.5}$, there is a small island of low  Alfv\'en speed, and that this is  zero at the X-point itself. Figure \ref{Figure2}c shows a colour contour of $v_A(0,y,z)$ along in the $yz-$plane at $x=0$. Note that in this plane, $v_A(0,y,z) = | {\bf{B}}_0(0,y,z)| =  | B_x(0,y,z)| =  | B_x(0,r)|$ as per equation (\ref{X-line_equation_2}). As in Figure \ref{Figure1}c, the blue line indicates the X-line. The contour is  colour coded in the same way as for Figure \ref{Figure2}b. {{The Alfv\'en speed is identically zero along the X-line, denoted by  $r^2=y^2+z^2=2a^2$.}} Figure \ref{Figure2}c can be further understood by Figure  \ref{Figure2}d which shows a plot of  $|B_x(0,r)|=v_A(0,y,z)$. Here, the green, yellow and orange colours correspond to those of Figure  \ref{Figure2}b and Figure  \ref{Figure2}c. However, note that here  black represents  $|B_x(0,r)| \le 0.3$.

%%%%%%%%%%%%%%%%%%%%%%%%%%%%%%%%%%%%%%%%%%%%%%%%%%%%%%%%%%%%%%%%%%%%%%%%%%%%%%%%%%%%%%%%%%%%%%%%%%%%%%

\section{WKB approximation}\label{WKBAPPROXIMATION}

In this paper, we will be looking for WKB solutions (see e.g. Bender \& Orszag \cite{Bender1978}) of the form:
\begin{eqnarray}
{{\mathbf{v}}_1} = {\bf{{\mathcal{V}}}} {\rm{e}}^{{\rm{i}} \phi (x,\:y,\:z,\:t) }  \label{WKB}
\end{eqnarray}
where ${\bf{{\mathcal{V}}}}$ is a constant vector. In addition, we define $\omega = \partial \phi / \partial t$ as the angular frequency and ${\bf{k}} = \nabla \phi=\left(p,q,r\right)$ as the wavevector. Note that $\phi$ and its  derivatives are considered to be the large parameters in our system.

One of the difficulties associated with three-dimensional MHD wave propagation is distinguishing between the three different wave types, {{i.e.}} between the fast and slow magnetoacoustic waves and the Alfv\'en wave. To aid us in our interpretation, we now define a coordinate system $( {\bf B}_{0},  {\bf k},  {\bf B}_{0} \times {\bf k}$) where ${\bf{k}}$ is our wavevector as defined above. This coordinate system fully describes all three directions in space when ${\bf B}_{0}$ and  ${\bf {k}}$ are not parallel to each other, {{i.e.}} $                       {\bf {k}} \neq \lambda       {\bf B}_{0} $, where $\lambda$ is some constant of proportionality. In $\S\ref{SEC:FAST}$, we will consider the fast wave solution. The Alfv\'en wave solution is considered in Appendix \ref{appendixB}. In $\S\ref{SEC:FAST}$, we will proceed assuming  $  {\bf {k}} \neq \lambda       {\bf B}_{0}        $. The scenario where  $   {\bf {k}} = \lambda       {\bf B}_{0}          $ is looked at in Appendix \ref{appendixA}. In fact, the work described in $\S\ref{SEC:FAST}$  is also valid for  ${\bf k} = \lambda{\bf {B}}_0$ with the consequence that the solution is degenerate, {{i.e.}} the waves recovered are identical and so the fast wave ($\S\ref{SEC:FAST}$) cannot be distinguished from the Alfv\'en wave  when  ${\bf k}  \propto {\bf {B}}_0$.

We now substitute ${{\mathbf{v}}_1} = {\bf{{\mathcal{V}}}} {\rm{e}}^{{\rm{i}} \phi (x,\:y,\:z,\:t) }$  into equation (\ref{10}) and make  the WKB approximation such that $\phi \gg 1$. Taking the scalar product with ${\bf{B}}_0$, $\bf{k}$ {{and}} ${{\bf{B}}_0} \times {{\bf{k}}}$  gives three velocity components which in matrix form are:
\begin{eqnarray*}
\left[ \begin{array}{ccc}
\omega^2 & 0  & 0 \\
\left( { {\bf{B}}_0 \cdot  {\bf{k}} } \right) \left|    {\bf{k}}   \right|^2  & \; \; \omega^2 -  \left|    {\bf{B}}_0   \right|^2       \left|    {\bf{k}}    \right|^2  & 0 \\
0 & 0 &  \omega^2  - {\left( {\bf{B}}_0 \cdot   {\bf{k}}   \right)^2}\end{array}\right]
 \left( \begin{array}{c}
{{{\bf v}_1}}\cdot  {\bf B}_{0} \\
{{{\bf v}_1}}\cdot    {\bf{k}}  \\
{{{\bf v}_1}}\cdot {  {\bf B}_{0}\times{    {\bf{k}}    }}\end{array}\right)  \nonumber \\
=
 \left (\begin{array}{c}
0 \\
0 \\
0 \end{array}\right)\;.
\end{eqnarray*}
In order to {{avoid the}} trivial solution, the matrix of these three coupled equations must have zero determinant. Thus, setting the determinant equal to zero gives:
\begin{eqnarray}
&& \mathcal{F}\left( \phi, \omega, t,  {\bf{B}}_0, {\bf{k}} \right)    \nonumber \\
&&=\omega^{2} \left( \omega^{2} - {\left|{\bf B}_0 \right|^2 \left|    {\bf{k}} \right|^2 } \right)  \left( \omega^{2}-\left( {\bf{B}}_0 \cdot {\bf{k}} \right) ^{2} \right) =0 \;, \label{F1}
\end{eqnarray}
where $\mathcal{F}$ is a first-order, non-linear partial differential equation. Equation (\ref{F1}) has two solutions, corresponding to two different MHD wave types: the fast magnetoacoustic wave and  the Alfv\'en wave. Note that in general there are three wave solutions, but the slow wave has vanished under the $\beta=0$ cold plasma approximation. We also do not consider the $\omega=0$ trivial solution. 

%%%%%%%%%%%%%%%%%%%%%%%%%%%%%%%%

The case where the two roots of equation (\ref{F1}) are the same is examined in Appendix \ref{appendixA}.

%%%%%%%%%%%%%%%%%%%%%%%%%%%%%%%%%%%%%%%%%%%%%%%%%%%%%%%%%%%%%%%%%%%%%%%%%%%%%%%%%%%%%%%%%%%%%%%%%%%%%%

\subsection{Fast wave solution}\label{SEC:FAST}

Let us consider the fast wave solution, and hence we assume $\omega^2 \neq \left( {\bf{B}}_0 \cdot {\bf{k}} \right) ^{2}$. Thus, equation (\ref{F1}) simplifies to:
\begin{eqnarray}
&&{\mathcal{F}} \left( \phi, \omega, t, {\bf{B}}_0,  {\bf{k}} \right) = \omega^{2} -  \left| {\bf{B}}_0   \right| ^2  \left|   {\bf{k}} \right| ^2   \nonumber \\
&&\Rightarrow {\frac{1}{2}} \left[ \omega^{2} - \left(B_x^2+B_y^2+ B_z^2\right) \left( p^2+q^2+r^2\right) \right]=0  \;,\; \label{fastEquation}
\end{eqnarray}
where we have introduced ${1}/{2}$ to simplify the equations later on. We can now use Charpit's method (e.g. see Evans \cite{Evans1999}) to solve this first-order partial differential equation, where we assume our variables depend upon some independent parameter $s$ in characteristic space.  Charpit's method replaces a first-order {{partial}} differential equation with a set of characteristics that are a system of  first-order {{ordinary}} differential equations. Here, Charpit's equations take the form:
\begin{eqnarray*}
\frac{ {\rm{d}} \phi}{{\rm{d}} s} &=& \left( {{\omega}}\frac{\partial} {\partial {{\omega}}}    +    {\bf{k}}\cdot \frac{\partial} {\partial {\bf{k}}} \right) \mathcal{F}\;\;, \quad\frac{{\rm{d}} {{t}}} {{\rm{d}} s} =\frac{\partial} {\partial {{\omega}}} \mathcal{F}  \;\;,\quad \frac{{\rm{d}} {\bf{x}}} {{\rm{d}} s} =\frac{\partial} {\partial {\bf{k}}} \mathcal{F}     \;\;,\\
 \frac{{\rm{d}} {{\omega}}}{{\rm{d}} s}&=& -\left( \frac{\partial} {\partial {{t}}} + {{\omega}} \frac{\partial} {\partial \phi}  \right) \mathcal{F}\;\;,\quad  \frac{{\rm{d}} {\bf{k}}}{{\rm{d}} s}= -\left( \frac{\partial} {\partial {\bf{x}}} + {\bf{k}} \frac{\partial} {\partial \phi}  \right) \mathcal{F}\;\;,
\end{eqnarray*}
where as previously defined ${\bf{k}}=(p,q,r)= \nabla \phi$ and ${\bf{x}}=(x,y,z)$.  These ordinary differential equations are subject to the initial conditions $\phi=\phi_0(s=0)$, $x=x_0(s=0)$,  $y=y_0(s=0)$, $z=z_0(s=0)$,   $t=t_0(s=0)$, $p=p_0(s=0)$, $q=q_0(s=0)$, $r=r_0(s=0)$ and $\omega=\omega_0(s=0)$  and are solved numerically using a fourth-order Runge-Kutta method.

Note that there are no boundary conditions in the traditional sense: the variables are solved using Charpit's method (essentially a variation on the method of characteristics) and the resulting characteristics are only dependent upon initial position $\left( x_0, y_0, z_0, t_0 \right)$ and the distance travelled along the characteristic, $s$. Thus, only initial conditions are required and no  boundary conditions are imposed. The fact that  WKB solutions are independent of  boundary conditions is actually an advantage over traditional numerical simulations where the choice of boundary conditions can play a significant role. In this paper, we have chosen to illustrate our results in the domain $-1 \le x \le 1$, $-1 \le y \le 1$, $0 \le z \le 2$, and this choice is arbitrary.

For the fast wave solution and  equation (\ref{fastEquation}), Charpit's equations are:
\begin{eqnarray}
{{\rm{d}}\phi \over {\rm{d}}s}&=&0 \;, \quad {{\rm{d}}t \over {\rm{d}}s}=\omega \;,\quad {{\rm{d}} \omega \over {\rm{d}}s}=0 \;,  \nonumber \\
\frac {dx}{ds} &=& - p  \left| {\bf{B}}_0   \right| ^2  \;  ,  \quad    \frac {dy}{ds} = - q  \left| {\bf{B}}_0   \right| ^2  \;  ,  \quad   \frac {dz}{ds} = - r  \left| {\bf{B}}_0   \right| ^2  \;  ,  \nonumber \\
\frac {dp}{ds} &=& \left( B_x \; \frac{\partial B_x}{\partial x} + B_y\frac{\partial B_y}{\partial x}  + B_z \; \frac{\partial B_z}{\partial x}  \right)\; \left|   {\bf{k}} \right| ^2  \;, \nonumber\\
\frac {dq}{ds} &=& \left( B_x \; \frac{\partial B_x}{\partial y} + B_y \; \frac{\partial B_y}{\partial y} + B_z \; \frac{\partial B_z}{\partial y} \right)\; \left|   {\bf{k}} \right| ^2  \;, \nonumber\\
\frac {dr}{ds} &=& \left( B_x \; \frac{\partial B_x}{\partial z} + B_y \; \frac{\partial B_y}{\partial z} + B_z \; \frac{\partial B_z}{\partial z} \right)\; \left|   {\bf{k}} \right| ^2  \;, \label{fast_dipole_characteristics}
\end{eqnarray}
where   $B_x$, $B_y$ and $B_z$ are the components of our equilibrium field, $\left| {\bf{B}}_0   \right| ^2 = B_x^2+B_y^2+B_z^2$, $\omega$ is the angular frequency of our wave, $s$ is the parameter along the characteristic,   $p=\frac {\partial \phi} {\partial x}$, $q=\frac {\partial \phi} {\partial y}$, $r=\frac {\partial \phi} {\partial z}$ and  $\left|   {\bf{k}} \right| ^2 = p^2+q^2+r^2$. We note that $\phi={\rm{constant}}=\phi_0$ and $\omega={\rm{constant}}=\omega_0$, i.e. constant angular frequency. In addition, $t=\omega s+t_0$ where we arbitrarily set $t_0=0$, which corresponds to the leading edge of the wave  starting at $t=0$ when $s=0$.  The other six ordinary differential equations are solved numerically using a fourth-order Runge-Kutta method.

%%%%%%%%%%%%%%%%%%%%%%%%%%%%%%%%%%%%

\subsection{Planar fast wave launched from $z_0=0.2$}\label{sub1}

We now solve equations (\ref{fast_dipole_characteristics}) subject to the initial conditions:
\begin{eqnarray}
\phi_0 &=&0 \;,\;\; \omega_0=2\pi \;,\;\; -1\leq x_0 \leq 1 \;,\;\;  -1 \leq y_0 \leq 1 \;, \nonumber \\
z_0 &=& 0.2 \;,\; p_0 = 0 \;,\; q_0 = 0  \;,\; r_0 = - \omega_0 /  {\left| {{\bf{B}}_0}  \left(x_0,y_0,z_0\right)  \right|  }\;,\label{chad}
\end{eqnarray}
where we have chosen arbitrarily $\omega_0 =2\pi$ and $\phi_0=0$. These initial conditions correspond to a planar fast wave starting at $z=z_0$ and  propagating in the direction of increasing $z$. {{From a modelling viewpoint, this choice of initial condition is intended to mimic a disturbance initiated at the \lq{photosphere}\rq{} of $z=z_0$.}}

Note that our choice of a magnetic dipole configuration for the equilibrium magnetic field has two singularities in the field at $(x,y,z)=(\pm a,0,0)$ and hence $v_A \rightarrow \infty$ at these points. Thus, if we were to start our planar wave at $z=0$ in the $xy-$plane, it would encounter this extreme speed differential. Thus, we generate our waves not at $z=0$ but at $z=z_0$, where $z_0$ is small. This choice still starts the waves in a region of strongly varying Alfv\'en speed, and so this choice results in very little loss of insight into the system. In this paper, we choose $z_0=0.2$.

%%%%%%%%%%%%%%%%%%%%%%%%%%%%%%%%%%%%%%%%%%%%%%%%%%%%%%%%%%%%%%%%%%%%%%

\begin{figure}[t]
\begin{center}
\includegraphics[height=6cm]{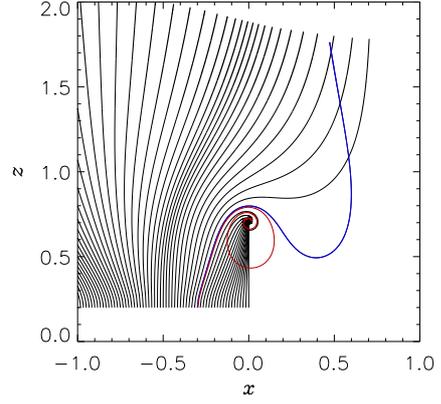}
\caption{Particle paths for starting points of $-1 \leq x_0 \leq 0$  set at intervals of $0.01$. The coloured lines represents the particle paths for starting points of $x_0 =-0.3$ (blue) and $x_0=-0.298$ (red) respectively.}
\label{Figure4}
\end{center}
\end{figure}

%%%%%%%%%%%%%%%%%%%%%%%%%%%%%%%%%%%%%%%%%%%%%%%%%%%%%%%%%%%%%%%%%%%%%%

\begin{figure*}[t]
\begin{center}
\includegraphics[height=16.1cm]{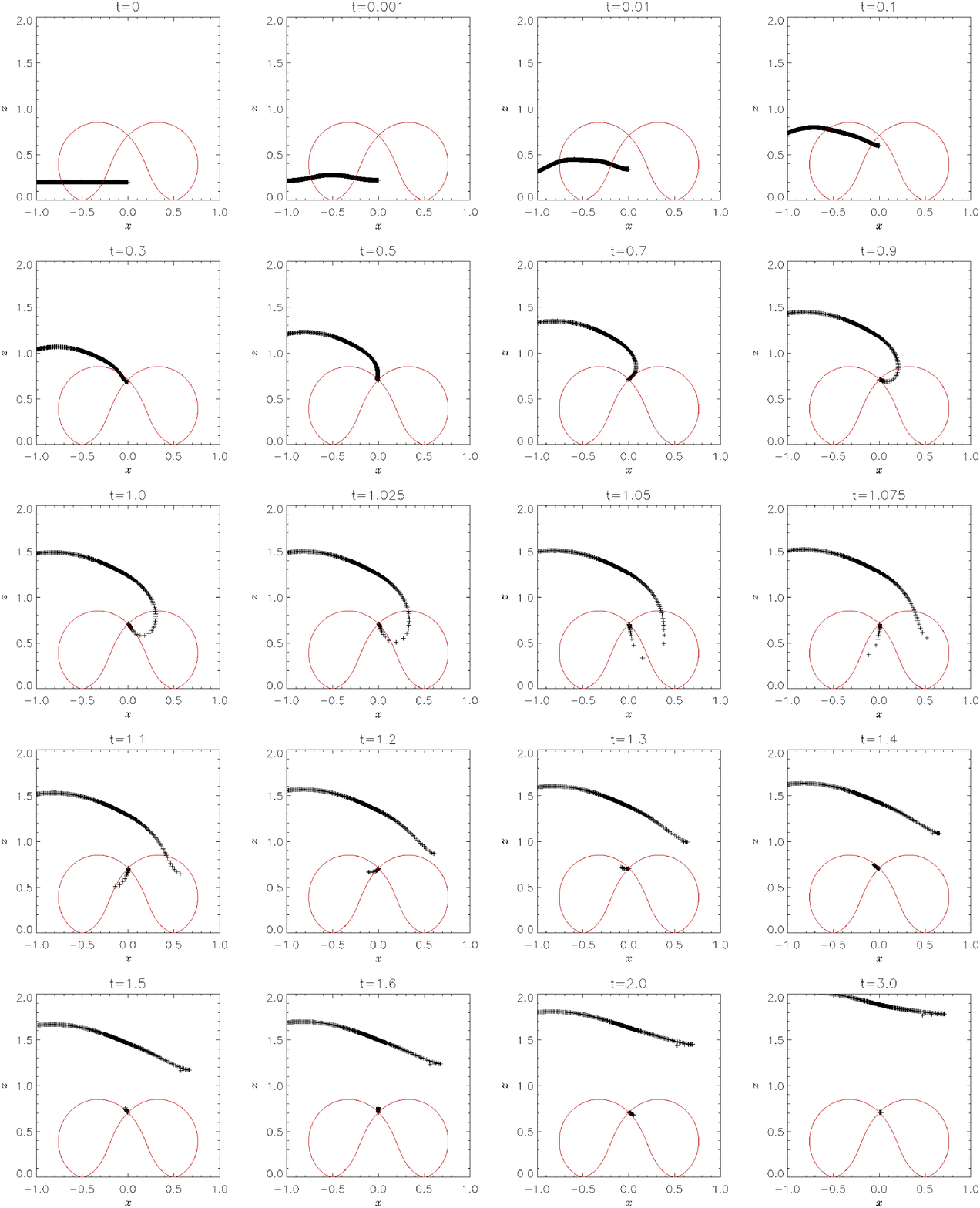}
\caption{Solution of constant values of $\phi$ for WKB approximation of a fast wave generated on lower boundary for $-1 \leq x_0 \leq 0 $, $y_0=0$, $z_0=0.2$ and its resultant propagation in the $xz-$plane at various times. Displayed times have been chosen to best illustrate evolution so  time between frames is not necessarily uniform. The wavefront  consists of crosses from the WKB wave solution, to better illustrate the evolution. The red separatrices in the $xz-$plane are also shown to provide context.}
\label{Figure3}
\end{center}
\end{figure*}

%%%%%%%%%%%%%%%%%%%%%%%%%%%%%%%%%%%%%%%%%%%%%%%%%%%%%%%%%%%%%%%%%%%%%%

\begin{figure}[t]
\begin{center}
\includegraphics[height=6cm]{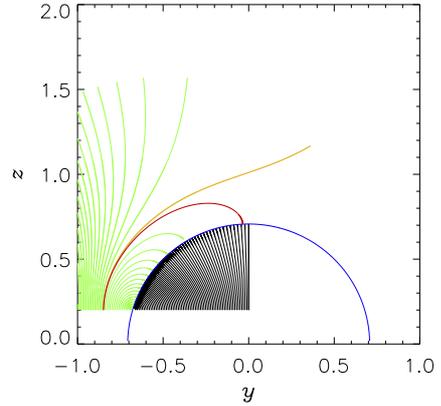}
\caption{Particle paths for starting points of $x_0=0$, $-1 \leq y_0 \leq 0$ and $z_0=0.2$. The X-line is indicated in blue. The lines for $- 1 \leq y_0 <  -0.6782$ have been coloured green to distinguish them from the lines  $- 0.6782 < y_0 \leq 0$  which are black. The orange and red lines represents the particle paths for a starting point of $y_0 =-0.85$  and $x_0=-0.848$, respectively.} 
\label{Figure6}
\end{center}
\end{figure}

%%%%%%%%%%%%%%%%%%%%%%%%%%%%%%%%%%%%%%%%%%%%%%%%%%%%

\begin{figure*}[t]
\begin{center}
\includegraphics[height=14cm]{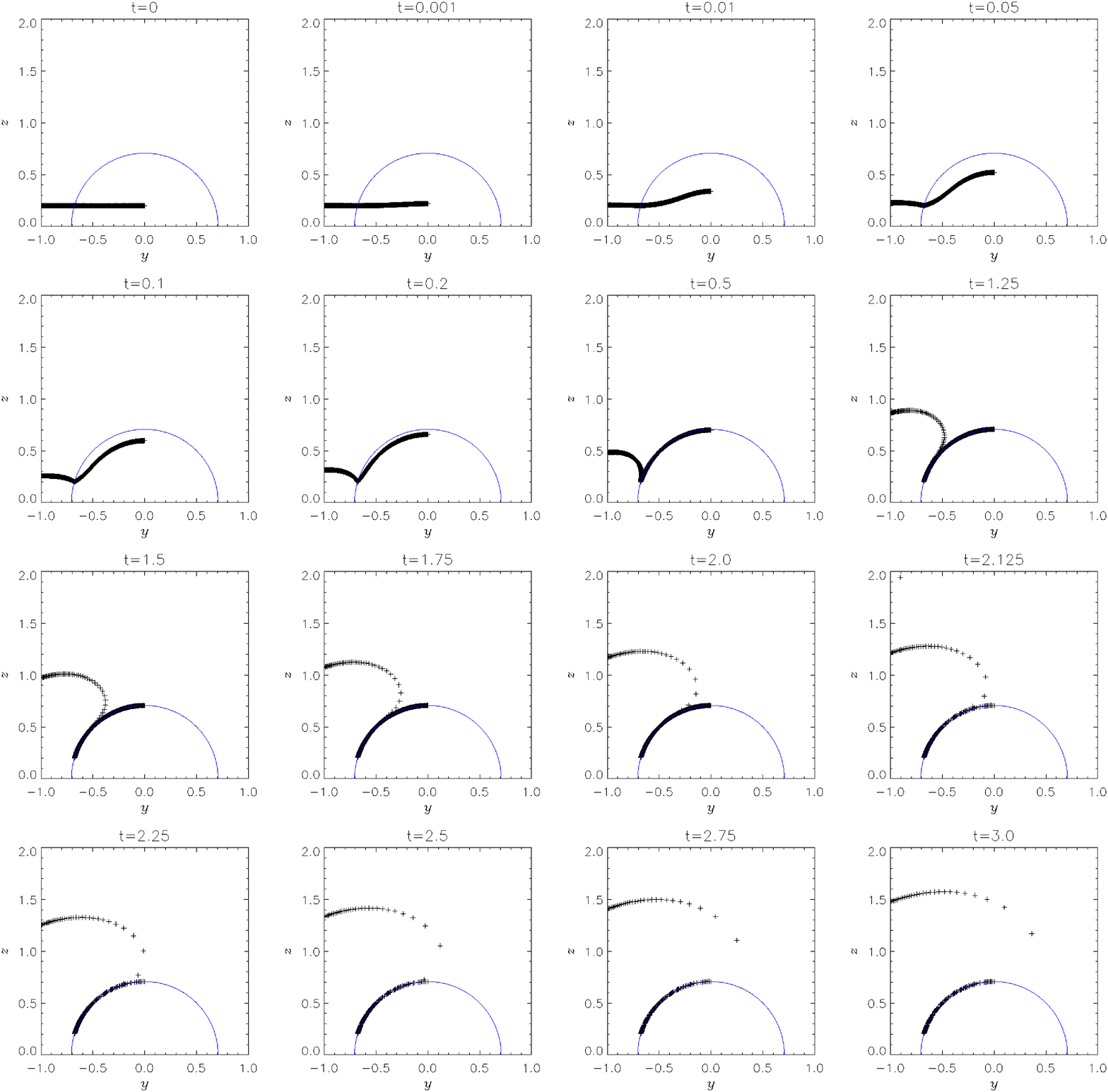}
\caption{Solution of constant values of $\phi$ for WKB approximation of a fast wave generated on lower boundary for $x_0=0$, $-1 \leq y_0 \leq 0 $, $z_0=0.2$ and its resultant propagation in the $yz-$plane at various times.  Displayed times have been chosen to best illustrate evolution so  time between frames is not necessarily uniform. The wavefront  consists of crosses from the WKB wave solution, to better illustrate the evolution.  The blue line indicates the location of the X-line.}
\label{Figure5}
\end{center}
\end{figure*}

%%%%%%%%%%%%%%%%%%%%%%%%%%%%%%%%%%%%%%%%%%%%%%%%%%%%

\begin{figure*}[t]
\begin{center}
\includegraphics[width=\textwidth]{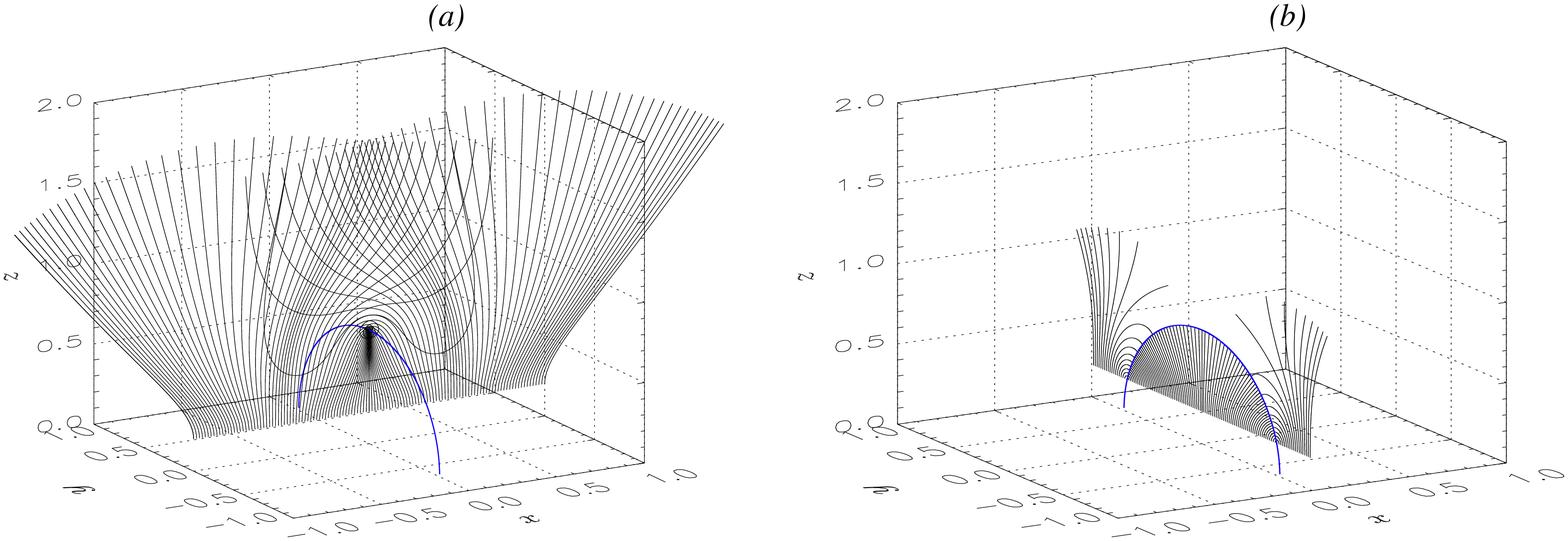}
\caption{Particle paths for individual elements generated along $(a)$  $-1 \leq x_0 \leq 1$, $y_0=0$, $z_0=0.2$ and $(b)$ along $x_0=0$,  $-1 \leq y_0 \leq 1$, $z_0=0.2$. The blue line indicates the location of the X-line.}
\label{figure7}
\end{center}
\end{figure*}

%%%%%%%%%%%%%%%%%%%%%%%%%%%%%%%%%%%%%%%%%%%%%%%%%%%%

\section{Results}\label{sec:results}

\subsection{Wave behaviour in the $xz-$plane along $y=0$}\label{subsec:1}

We now look at the behaviour and evolution of the fast wave solution in the neighbourhood of our two dipoles. The WKB solution permits us two approaches to investigating the wave behaviour: we can visualise the particle paths or ray paths of individual wave elements  generated at specific start points (this can be seen in Figure \ref{Figure4}) or we can plot surfaces of constant $\phi$ at various times which can be thought of as defining the location of the wavefront (this can be seen in Figure \ref{Figure3}).

Figure \ref{Figure4}  plots the particle paths of individual elements from the initially planar wave.  To best illustrate the behaviour, we plot the particle paths for starting points of $-1 \leq x_0 \leq 0$ set at intervals of $0.01$. The system is symmetric and so the behaviour for  $0 \leq x_0 \leq 1$ can also be understood under the transformation $-x \rightarrow x$.  We see that the lines for  $x_0 \leq -0.5$ do not appear to be influenced greatly by the X-point and simply propagate in the direction of increasing $z$ at varying angles. However, the particle paths for starting points of $-0.5 \le x_0 < -0.3$ are influenced heavily by the X-point, but only in so much as to deflect the ray path. The blue line represents the particle path for a starting point of $x_0 =-0.3$ and again, though the ray path is influenced significantly by the topology, namely refracted towards the X-point, i.e. a region of lower Alfv\'en speed, and then refracted away from the dipole loci close to $x=0.5$, i.e. a region of high Alfv\'en speed, this individual element still eventually escapes the magnetic field configuration. We call this starting point $x_{\rm{critical}}=-0.3$. For starting points greater than $x_{\rm{critical}}=-0.3$, the particle paths spiral towards the X-point and are ultimately trapped there. The red line represents the particle path for a starting point of $x_0 =-0.298$ and clearly shows this spiralling effect, i.e. the particle is refracted into the region of low Alfv\'en speed around the X-point and wraps around it. Thus, there are  two types of behaviour: either the ray paths are trapped by the X-point or ultimately escape, and there is a critical starting point that divides these two types of behaviour. For the system studied here, and due to symmetry,  this is  $x_0 = x_{\rm{critical}}= \pm 0.3$.

%%%%%%%%%%%%%%%%%%%%%%%%%%%%%%%%%%%%%%%%%%%%%%%%%%%%%%%%%%%%%%%%%%%%%%

Let us now consider the propagation of a wavefront as opposed to the particle paths of individual elements. Figure \ref{Figure3} shows several plots of constant $\phi$ at various times, which can be thought of as defining the position of the wavefront. Since $t=\omega s$, each time also corresponds to a different value of the parameter $s$, which quantifies the  distance travelled along the characteristic curve. Each individual element is thus fully described by its initial starting position $(x_0,y_0, z_0)$, where $z_0$ is fixed, and its evolution according to $s$. In Figure \ref{Figure3}, {{we have plotted each individual element as a cross to better illustrate}} the  wave stretching and splitting effects that occur. The evolution of the wavefront is shown for $0 \le t \le 3$.  Displayed times have been chosen to best illustrate the evolution, e.g. more subfigures are presented to detail the splitting between $1 \le t \le 1.1$ and so the time between frames is not uniform. We present the results for a wavefront generated on  $-1 \leq x_0 \leq 0$ so as to best illustrate the subsequent behaviour.

We find that the fast wave wavefront starting between $-1 \leq x_0 \leq 0$ propagates upwards (in the direction of increasing $z$) from the lower boundary $z_0$, but not all parts rise at the same speed. The central part of the wave rises much faster, with the maximum occurring over $x=-0.5$. This is due to the high Alfv\'en speed localised in that area as seen in Figure \ref{Figure2}b. This inhomogeneous Alfv\'en speed profile deforms the wave from its original planar form, since each individual part of the wavefront propagates with its own local (Alfv\'en) speed. Part of the wave pulse also approaches the X-point, i.e. the part of wave between $x_{\rm{critical}} < x \le 0 $, and this part gets caught around the X-point. The subsequent evolution now takes two different forms: with part of the wave being trapped by the X-point and part being deformed by the varying Alfv\'en speed but, ultimately, escaping and propagating in the direction of increasing $z$. Thus, there is a critical starting point that divides these two types of behaviour,  $x_{\rm{critical}}$, in agreement with Figure \ref{Figure4}.

Let us first consider the part of the wave captured by the X-point, i.e.  the part of wave generated between $x_{\rm{critical}} < x \le 0$ at $y=0$ and $z=z_0$. {{This part of the wave propagates upwards and begins to wrap around the X-point, due to the variation in Alfv\'en speed as seen in Figure \ref{Figure2}b. Ultimately, the wave wraps itself around the X-point. As seen in Figure \ref{Figure2}d, the Alfv\'en speed is zero at the X-point  and so the fast wave cannot cross this point}}. Consequently the X-point acts as a focus for the refraction effect.

Let us now consider the part of the wave that escapes an ultimate fate of ending up at the X-point, i.e. the  wave generated between $-1 \le x \le x_{\rm{critical}}$ at $y=0$ and $z=z_0$. This part of the wave continues to propagate upwards  and spread out.  This part of the wave propagates at a slower speed than at earlier times, again due to the change in the strength of the local Alfv\'en speed. Once above the magnetic skeleton, the wave continues to rise and spread out:  the wave is no longer influenced by the X-point and  it has escaped the refraction effect. Ultimately this part of the wave leaves the presented domain completely.

However, since part of the wave is wrapping around the X-point and a second part is rising away, the wavefront is being pulled in two different directions. This can be seen {{in  Figure \ref{Figure3} at times $1 \le t \le 1.1$.}} Turning our attention to the behaviour to the right of the X-point, we see that the wave continues to spread out: part of it propagates towards the top right corner and part of it is hooked around the X-point. {{We see the wave is stretched between these two destinations. Part of the wave then ultimately wraps around the X-point and the other part propagates away from the dipole region. This gives the appearance of the wave splitting, however this is only due to our (discretised) plotting of the wavefront as individual crosses. What has actually occurred is extreme stretching of the wavefront. The extreme stretching occurs when the wave enters the right-hand region of high Alfv\'en speed, i.e. large $v_A$ around $x = 0.5$. 

When this extreme stretching occurs, the length scales across (perpendicular) to the stretching will rapidly decrease and so the local gradients will increase. Hence, if even a small amount of resistivity was included in our system, ohmic heating will act to extract the energy from this location. This would lead to a genuine splitting of the wavefront.

}}

Thus, a fast wave generated between $-1 \leq x \leq 0$ along $y=0$ and $z=z_0$ propagates  into the magnetic region unevenly due to the inhomogeneity in Alfv\'en speed profile. Part of the wave experiences a refraction effect, of various magnitudes, due to the non-uniform local Alfv\'en speed, but ultimately spreads out and propagates away from the dipolar regions. However, part of the wave is caught by the X-point, refracted into it and  accumulates eventually at the X-point. Thus, {{there is a critical starting point that divides these two types of behaviour}} and investigation of the particle paths shows that for the system studied here this occurs at $x_0 = x_{\rm{critical}}= \pm 0.3$.

%%%%%%%%%%%%%%%%%%%%%%%%%%%%%%%%%%%%%%%%%%%%%%%%%%%%%%%%%%%%%%%%%%%%%%

We can also use the WKB approximation  to plot a solution for a wave generated at $-1 \leq x_0 \leq 1 $, $y_0=0$, $z_0=0.2$. This can be seen in  Figure \ref{Figure_appendix_1} in Appendix \ref{appendixC}. As before, each wavefront consists of many tiny crosses. Of course, the system is symmetric across $x=0$ and so the explanation of the behaviour is the same as that above.

%%%%%%%%%%%%%%%%%%%%%%%%%%%%%%%%%%%%%%%%%%%%%%%%%%%%%%%%%%%%%%%%%%%%%%%%%%%%%%%%%%%%%%%%%%%%%%%%%%%%%%%%%%%%%%%%%%%%

\begin{figure*}[t]
\begin{center}
\includegraphics[height=11cm]{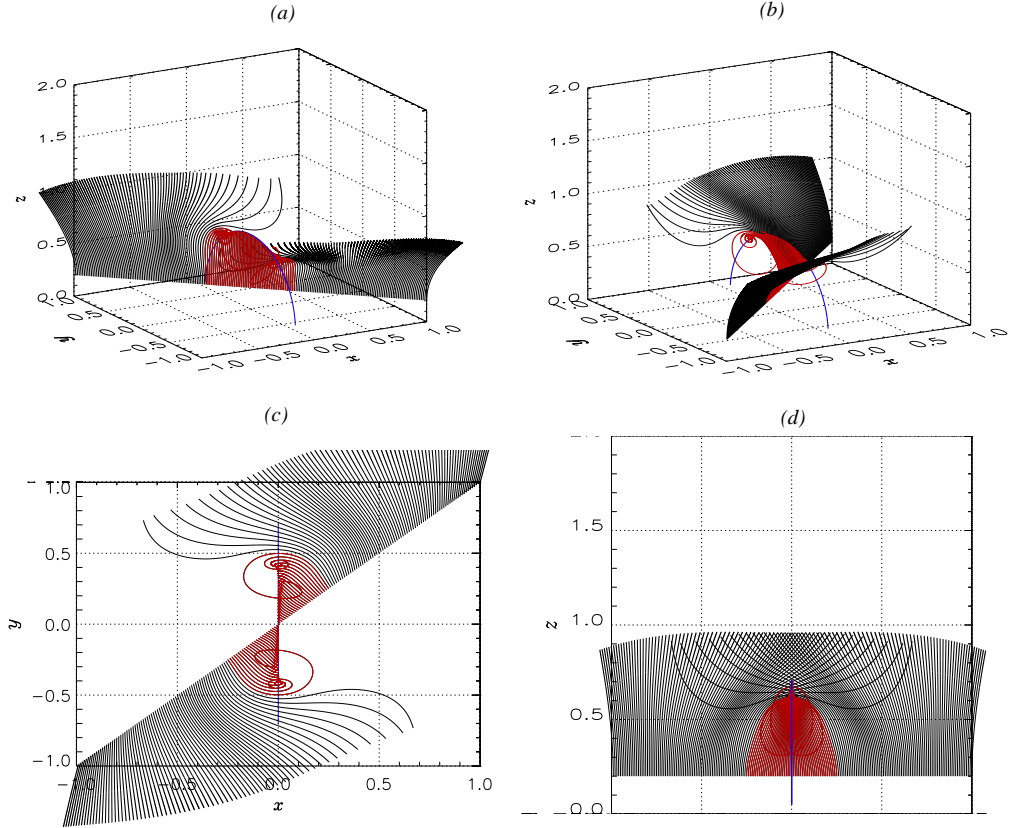}
\caption{Particle  paths for individual elements that  start along $(a)$ $y_0=-x_0$. $(b)$  $y_0=x_0$. $(c)$    $y_0=x_0$  looking down on the $xy-$plane, i.e. there is a line-of-sight effect along the $z-$direction and X-line is viewed from above. $(d)$  $y_0=x_0$  looking down on the $xz-$plane, i.e. there is a line-of-sight effect along the $y-$direction  and  X-line is viewed end-on. Blue line denotes the X-line. Ray paths for $|x_0| < 0.252$ are coloured red and are trapped by the X-line.}
\label{ffigure8}
\end{center}
\end{figure*}

%%%%%%%%%%%%%%%%%%%%%%%%%%%%%%%%%%%%%%%%%%%%%%%%%%%%%%%%%%%%%%%%%%%%%%%%%%%%%%%%%%%%%%%%%%%%%%%%%%%%%%%%%%%%%%%%%%%%%%%%%%%%%%%%%%%%%%%%%%%%%%%%%%%%%%%%%%%%%%%%%%%%%%%%%%%%%%%%%%%%%%%%%%%%%%%%%%%%%%%%%%%%%%%%%%%%%%%%

\begin{figure*}[t]
\begin{center}
\includegraphics[height=11cm]{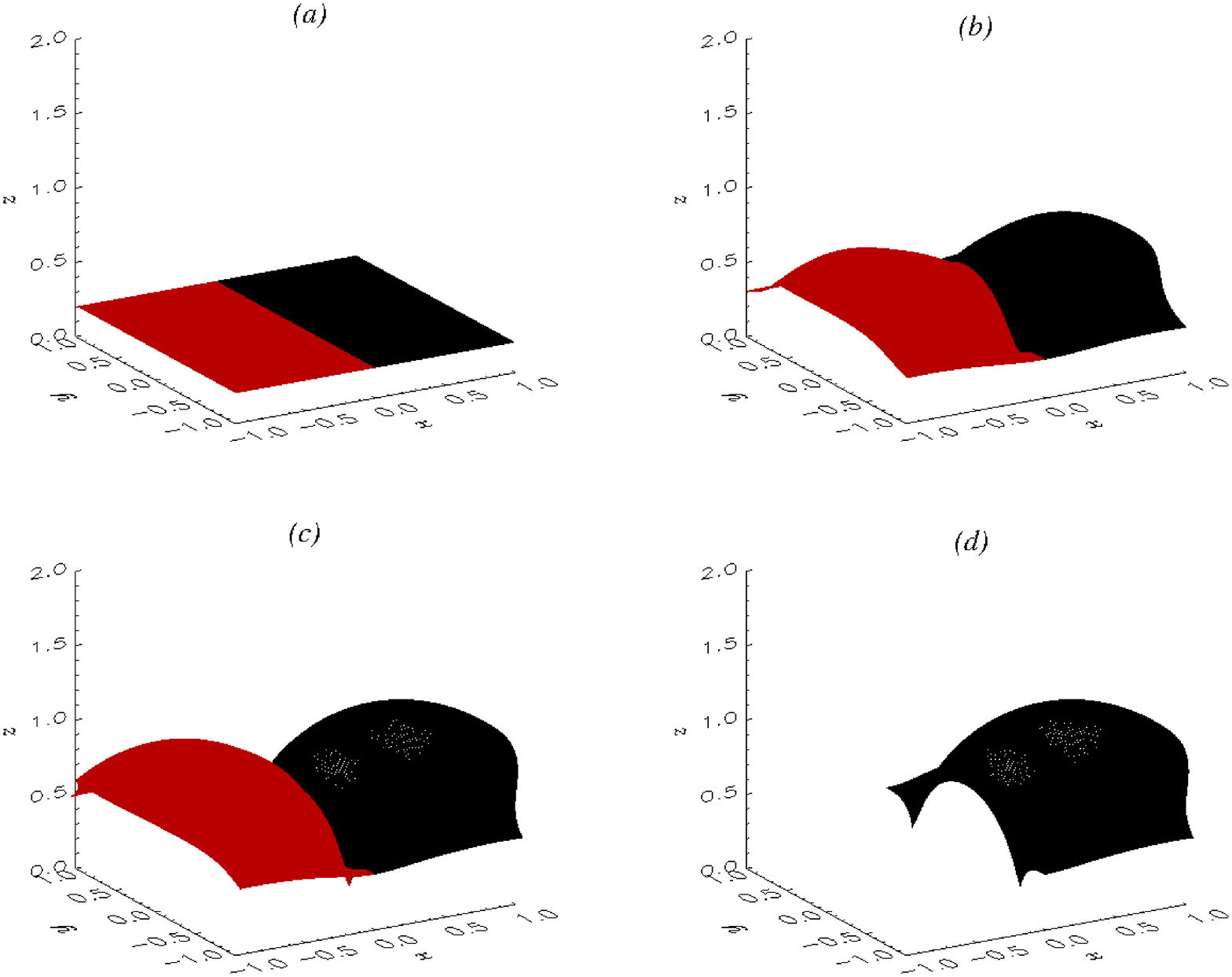}
\caption{Surfaces of  constant $\phi$ at  times $(a)$ $t=0$,  $(b)$ $t=0.1$ and $(c)$ $t=0.3$ for initially-planar wavefront generated on $-1 \leq x_0 \leq 1$,  $-1 \leq y_0 \leq 1$, $z_0=0.2$.  $-1 \leq x_0 \leq 0$ is coloured red and  $0 \leq x_0 \leq 1$ black to aid the reader in tracking the evolution. $(d)$ Same as $(c)$ but only for $0 \leq x_0 \leq 1$,  $-1 \leq y_0 \leq 1$, $z_0=0.2$ in order to highlight the behaviour along $x=0$.}
\label{dfigure9}
\end{center}
\end{figure*}

%%%%%%%%%%%%%%%%%%%%%%%%%%%%%%%%%%%%%%%%%%%%%%%%%%%%%%%%%%%%%%%%%%%%%%%%%%%%%%%%%%%%%%%%%%%%%%%%%%%%%%%%%%%%%%%%%%%%%%%%%%%%%%%%%%%%%%%%%%%%%%%%%%%%%%%%%%%%%%%%%%%%%%%%%%%%%%%%%%%%%%%%%%%%%%%%%%%%%%%%%%%%%%%%%%%%%%%%

\subsection{Wave behaviour in the $yz-$plane along $x=0$}\label{subsec:2}

We now look at the behaviour of the fast wave solution in the $x=0$, $yz-$plane, i.e. the plane that contains our X-line. This can be seen in Figure \ref{Figure6} which depicts the particle paths for starting points of $-1 \leq y_0 \leq 0$ set at intervals of $0.01$. Here, $x_0=0$ and $z_0=0.2$. The system is symmetric and so the behaviour for  $0 \leq y_0 \leq 1$ can also be understood under the transformation $-y \rightarrow y$.

For a wave generated at $z_0=0.2$, the straight line segment under the X-line is bounded by $y=\pm \sqrt{ 2a^2 - z_0^2} = \pm \sqrt{ 0.5-0.2^2}=\pm 0.6782$.  The rays from $- 1 \leq y_0 <  -0.6782$ have been coloured green to distinguish them from the lines  $- 0.6782 < y_0 \leq 0$  which are coloured black. We see that for  $ - 0.6782 < y_0 \leq 0$  the (black) generated ray paths propagate upwards from the lower boundary $z_0$ and all terminate at the X-line.  Here, the individual elements of the wave cannot cross the X-line due to the  zero  Alfv\'en speed along those locations and thus this is where the wave accumulates. This propagation can be understood by looking at the  Alfv\'en speed profile in Figure \ref{Figure2}d, which shows that the magnitude of the wave speed decreases as an individual element approaches the X-line, and is equal to zero at $r=\sqrt{0.5}=0.707$. This result, i.e. that fast waves accumulate along X-lines, is a new phenomenon which has not been reported in previous papers.

For $- 1 \leq  y_0 <  -0.6782$, the  (green) ray paths are deflected by the varying Alfv\'en speed profile and we observe two types of behaviour. In Figure \ref{Figure6}, the orange and red lines represents the particle paths from a starting point of $y_0 =-0.85$  and $y_0=-0.848$, respectively.  For $ y_0 \leq -0.85$, we see that the  ray paths are influenced heavily by the inhomogeneous Alfv\'en speed profile but ultimately escape the system. However, for  $ y_0 > -0.85$ the ray paths refract towards the X-line and terminate there.  Thus, {{as in $\S\ref{subsec:1}$}}, {{there is a critical starting point that divides these two types of behaviour}}. For the system studied here, and due to symmetry,  this is  $y_0 = y_{\rm{critical}}= \pm 0.85$.

Figure \ref{Figure5}  shows plots of constant $\phi$ at various times. The evolution of the wavefront is shown for $0 \le t \le 3$.  Displayed times have been chosen to best illustrate the evolution, e.g. more subfigures are presented to detail the splitting between $2 \le t \le 2.5$ and so the time between frames is not uniform. We present the results for a wavefront generated on  $-1 \leq y_0 \leq 0$ so as to best illustrate the subsequent behaviour.

We find that the fast wave solution starting between $- 0.6782  < y_0 \leq  0.0$, i.e. under the X-line, propagates upwards from the lower boundary $z_0$ and accumulates along the X-line. Note that elements generated at the X-line itself, i.e. $x_0=0$, $y_0=  0.6782$, $z_0=0.2$, have zero local Alfv\'en speed and so remain stationary.

For the fast wave solution starting between $-1 \leq y_0 < -0.6782$, we find that elements of the wavefront generated on $ -0.85 < y_0 < - 0.6782$ are refracted into the X-line, whereas elements generated $-1 \leq y_0 \leq -0.85$ ultimately escape the X-line, although their propagation is modified by the inhomogeneous Alfv\'en speed profile. From  Figure \ref{Figure6}, this corresponds to  $y_{\rm{critical}}= -0.85$.

%%%%%%%%%%%%%%%%%%%%%%%%%%%%%%%%%%%%%%%%%%%%%%%%%%%%%%%%%%%%%%%%%%%%%%

We can also use the WKB approximation  to plot a solution for a wave generated at $x_0=0$, $-1 \leq y_0 \leq 1 $ and $z_0=0.2$. This can be seen in Figure \ref{Figure_appendix_2} in Appendix \ref{appendixC}. As before, each wavefront consists of many tiny crosses. Of course, the system is symmetric across $y=0$ and so the explanation of the behaviour is the same as that above.

%%%%%%%%%%%%%%%%%%%%%%%%%%%%%%%%%%%%%%%%%%%%%%%%%%%%%%%%%%%%%%%%%%%%%%

\subsection{Three-dimensional particle paths launched from $z_0=0.2$}\label{sub3}

We can also use our WKB solution to plot the three-dimensional particle  paths of individual fluid elements generated at $(x_0, y_0, z_0=0.2)$. In Figure \ref{figure7}a, we see the particle  paths for individual elements that begin at  starting points  of $-1 \leq x_0 \leq 1$ set at intervals of $0.01$ along $y_0=0$ and $z_0=0.2$. Thus, this is a comparison figure for Figure \ref{Figure4} and Figure  \ref{Figure_appendix_1}.  Figure \ref{figure7}b depicts the  particle  paths for individual elements that begin at  starting points  of $-1 \leq y_0 \leq 1$ set at intervals of $0.01$ along $x_0=0$ and $z_0=0.2$, i.e.   a comparison figure for Figure \ref{Figure6} and Figure \ref{Figure_appendix_2}.  In both, the blue line indicates the location of the X-line. The results above have shown that it is the X-line that plays a key role, rather than the separatrices. Hence, we do not plot the separatrices in our 3D figures.

Figure \ref{ffigure8}a and Figure \ref{ffigure8}b show the  particle  paths for individual elements that  start along the line $y_0=-x_0$ and $y_0=x_0$ respectively. We see that, as detailed in $\S\ref{subsec:1}$ and $\S\ref{subsec:2}$, there are two types of ray behaviour: rays can be trapped by the X-line and ultimately terminate there, or can eventually  escape the X-line, where the closer a ray gets to the X-line the stronger its deflection/modification by the local Alfv\'en speed profile.

Along the line $y_0=\pm x_0$, elements that are generated for $|x_0| < 0.252$ are coloured red and we see that these are all trapped by the X-line. Elements generated on $|x_0| \geq 0.252$ are coloured black and ultimately escape the X-line. Figure \ref{ffigure8}c and Figure \ref{ffigure8}d  show the same  particle  paths and colouring as  Figure \ref{ffigure8}b, i.e. for individual elements that  start along the line  $y_0=x_0$, but with a rotated perspective. Figure \ref{ffigure8}c shows the same particle paths looking down on the $xy-$plane, i.e. there is a line-of-sight effect along the $z-$direction and one is looking upon the X-line from above. Figure \ref{ffigure8}d shows the  same particle paths looking across  the $xz-$plane, i.e. there is a line-of-sight effect along the $y-$direction and one is looking at the X-line end-on. We see that it is the proximity to the X-line that entirely dictates the behaviour, in this case the critical value being $|x_0| = 0.252$.

%%%%%%%%%%%%%%%%%%%%%%%%%%%%%%%%%%%%%%%%%%%%%%%%%%%%%%%%%%%%

Finally, we can consider the propagation of an entire wavefront, as opposed to individual ray paths. Figure \ref{dfigure9}  shows surfaces of  constant $\phi$ at three particular times, showing the behaviour of the initially-planar wavefront that is generated on $-1 \leq x_0 \leq 1$,  $-1 \leq y_0 \leq 1$, $z_0=0.2$. We have coloured  $-1 \leq x_0 \leq 0$ red and  $0 \leq x_0 \leq 1$ black to aid the reader in tracking the wave behaviour. Figure \ref{dfigure9}a,  \ref{dfigure9}b and \ref{dfigure9}c show the wavefront at times $t=0$, $t=0.1$ and $t=0.3$ respectively. We see that the initially-planar wavefront propagates away from $z_0=0.2$ in the direction of increasing $z$. The wavefront is distorted due to the inhomogeneous Alfv\'en speed profile. From the results detailed above, we know that the behaviour after $t=0.3$ involves wrapping around the X-line and so the surfaces of constant $\phi$ become significantly distorted, i.e. wrapping back on themselves, and thus there is little extra information to be gained from looking at such figures for $t>0.3$. Hence, we only present the surfaces of constant $\phi$ at these early times. 

Figure \ref{dfigure9}d  shows the same surface as Figure \ref{dfigure9}c at $t=0.3$ but only for  the  wavefront  generated on $0 \leq x_0 \leq 1$,  $-1 \leq y_0 \leq 1$, $z_0=0.2$ (black). This removes the red surface and  allows us to see and highlight the behaviour  along $x=0$. As expected, the wavefront cannot cross the X-line and so it trapped there: one can clearly see the outline of the X-line along $x=0$ in Figure \ref{dfigure9}d.

%%%%%%%%%%%%%%%%%%%%%%%%%%%%%%%%%%%%%%%%%%%%%%%%%%%%

\section{Conclusion}\label{conclusion}

This paper describes an investigation into the behaviour of fast magnetoacoustic waves in the neighbourhood of two magnetic dipoles, under the assumptions of ideal and cold plasma. We have  demonstrated how the WKB approximation can be used to help solve the linearised MHD equations and we have utilised  Charpit's method and a Runge-Kutta numerical scheme during our investigation.

For the fast magnetoacoustic wave, we find that the wave speed is entirely dictated by the local equilibrium Alfv\'en speed profile. For individual elements generated on a lower plane, where here a wavefront was generated on the $xy-$plane at $z_0=0.2$, we find  that all parts of the wave experience a refraction effect, i.e. a deviation towards regions of lower Alfv\'en speed, and that the magnitude of the refraction is  different depending upon where a fluid element is in the magnetic field configuration. We find that there are two main types of wave behaviour: 
\begin{itemize}
\item{Individual fluid elements can be  trapped by the X-line, {{spiralling}} into the X-line due to the refraction effect. These individual elements terminate at the X-line. The wave speed decreases as an element approaches the X-line and the speed is identically zero at the X-line. Hence, it cannot be crossed.}
\item{Individual fluid elements can escape the system, where elements closer to the X-line have their ray paths modified to a greater extent that those {{farther}} away.}
\end{itemize}
Thus,  {\emph{there is a critical starting point that divides these two types of behaviour}}. We find that in the $xz-$plane along $y=0$, this  critical starting point is $x_0 = x_{\rm{critical}}= \pm 0.3$, and in the the $yz-$plane along $x=0$, this  critical starting point is $y_0 = y_{\rm{critical}}= \pm 0.85$. For starting positions along the lines $y_0 = \pm x_0$, it was found that the critical starting  point was $|x_0| = 0.252$.

We can also estimate the amount of wave energy trapped by the X-line. For the system studied here, the fraction captured by the X-line will depend upon the critical starting  point that divides the particle paths into those that spiral into the X-line and those that escape, as well as the overall length of the domain. In this paper we have set $-L \leq x \leq L$, $-L \leq y \leq L$  and $z=z_0=0.2L$, where $L$ is the length of our lower boundary and $L=1$ under our non-dimensionalisation. Thus along the line $x=0$, $y_{\rm{critical}}= \pm 0.85$ and so $85\%$ of the wave energy is trapped by the X-line, whereas  along $y=0$,  $x_{\rm{critical}}= \pm 0.3$ and so $30\%$ is trapped. Thus, more wave energy is trapped from the wave along the X-line than across it. Similarly, it was found that the critical starting  point along the lines $y_0 = \pm x_0$ was $|x_0| = 0.252$, which corresponds to $17.8 \%$ of the wave energy being trapped along a diagonal line of length $\sqrt{2}L$, or $25.2 \%$ being trapped along a radial line of $L=1$. Further investigation shows that there is a critical $\emph{area}$ surrounding the X-line, within which all wave elements and thus wave energy is trapped. This critical area corresponds to $0.618 \: L^2$ across  $-L \leq x \leq L$, $-L \leq y \leq L$  and $z=z_0=0.2L$, which corresponds to $15.5\%$ of the wave energy generated across an area $4\: L^2$ being trapped by the X-line. {{This  critical area is fixed for the magnetic topology}} and so the percentage trapped decreases as one increases the area of the initial wave considered.

%%%%%%%%%%%%%%%%%%%%%%%%%%%%%%%%%%%%%%%%%%%%%%%%%%%%%%%%%%%%

We have also limited our investigation to understanding the fast wave, but we could have also investigated the second  root of equation (\ref{F1}), i.e. the equations governing the Alfv\'en wave behaviour. To do so, we would assume  $\omega^2 \neq { | {\bf{B}}_0 |}^2 {| {\bf{k}} |} ^{2}$ and investigate the resultant equations. We have included such a derivation in Appendix \ref{appendixB} although a full investigation is outside the scope of this current paper.

%%%%%%%%%%%%%%%%%%%%%%%%%%%%%%%%%%%%%%%%%%%%%%%%%%%%%%%%%%%%

The 3D WKB technique  described in this paper can also be applied to other magnetic configurations and we hope that this paper has illustrated the potential of exploiting the  technique. In addition, it is possible to extend the work by dropping the cold plasma assumption. This will lead to a third root of equation (\ref{F1}) which will correspond to the behaviour of the slow magnetoacoustic wave. {{When the cold plasma assumption is dropped the fast wave speed will no longer be zero along the X-line, and thus the wave may pass through it. McLaughlin \& Hood (\cite{MH2006b}) investigated the behaviour of magnetoacoustic waves in a finite-$\beta$ plasma in the neighbourhood of a two-dimensional X-point. It was found that the fast wave could now pass through the X-point due to the non-zero sound speed and that a fraction of the incident fast wave was converted to a slow wave as the wave crossed the $\beta=1$ layer. }}

{{There are}} some caveats concerning the method presented here, i.e. if modellers wish to compare their work with a WKB approximation, it is essential to know the limitations of such a method.  Firstly, in linear 3D MHD, we would expect a coupling between the fast wave and the other  wave types due to the geometry. However, under the WKB approximation presented here, the wave sees the field as locally uniform  and so there is no coupling between the wave types. To include the coupling, one needs to include the next terms in the approximation, {{i.e.}} the work presented here only deals with the first-order terms of the WKB approximation. Secondly, note that the  work here is valid strictly for high-frequency waves, since we took $\phi$ and hence $\omega=\partial \phi / \partial t$ to be a large parameter in the system. The extension to low frequency waves is considered in Weinberg (\cite{Weinberg1962}).

{{In this paper}} we have found that the X-line acts as a focus for the refraction effect and that this {{refraction effect is a key feature of fast wave propagation}}. Since the Alfv\'en speed drops to identically zero at the X-line, mathematically the wave never reaches there, but physically the length scales (i.e. the distance between, say, the leading edge and trailing edge of a wave pulse and/or wave train) will rapidly decrease, indicating that all gradients, including  current density, will increase at this location.  In other words, {\emph{the fast wave, and thus all the fast wave energy, accumulates along the X-line}}. If even a small amount of resistivity was included in our system, ohmic heating will extract the energy from this location. {{ Thus, we deduce that {{X-lines will be specific locations of fast wave energy deposition and preferential heating}}. This highlights the importance of understanding the magnetic topology of a system and it is at these areas where preferential heating will occur. This paper specifically concerns itself with preferential heating at the X-line. However, it is important to note that these are not the only topological locations at which heat deposition is expected.}}

{{

Finally, we note that an X-line is a degenerate structure and its existence requires a special symmetry of the field, and any arbitrarily small perturbation to this symmetric configuration will lead to a magnetic topology without a true X-line. The resulting new topology may exhibit a non-zero component of ${\bf{B}}$ all along the original X-line, which may manifest as a quasi-separatrix layer, or as one or multiple null points. Should the symmetry be broken and the topology changed, we expect that (i) should a quasi-separatrix layer manifest, then we would still get the extreme stretching described in this paper, since the quasi-separatrix layer  would be a location of rapidly changing magnetic field connectivity, and hence all gradients, including  current density, may increase at these locations. (ii) Should null points appear, then we would expect to recover the results  of McLaughlin et al. (\cite{MFH2008}), who studied wave propagation around  3D null points.

}}

%%%%%%%%%%%%%%%%%%%%%%%%%%%%%%%%%%%%%%%%%%%%%%%%%%%%%%%%%%%%%%%%%%%%%%%%%%%%%%%%%%%%%%%%%%%%%%%%%%%%%%%%%%%%%%%%%%%%%%%%%%%%%%%%%%%%%%%%%%%%%%%

\begin{acknowledgement}
D.L. Spoors acknowledges an Undergraduate Research Bursary from the Royal Astronomical Society. The authors acknowledge IDL support provided by STFC. J.A. McLaughlin acknowledges generous support from the  Leverhulme Trust and this work was funded by a Leverhulme Trust Research Project Grant: RPG-2015-075. J.A. McLaughlin wishes to thank  Alan Hood for insightful discussions and helpful suggestions regarding this paper.

\end{acknowledgement}

%%%%%%%%%%%%%%%%%%%%%%%%%%%%%%%%%%%%%%%%%%%%%%%%%%%%%%%%%%%%%%%%%%%%%%%%%%%%%%%%%%%%%%%%%%%%%%%%%%%%%%%%%%%%%%%%%%%%%%%%%%%%%%%%%%%%%%%%%%%%%%%

%%%%%%%%%%%%%%%%%%%%%%%%%%%%%%%%%%%%%

\appendix

%%%%%%%%%%%%%%%%%%%%%%%%%%%%%%%%%%%%%%%%%%%%%%%%%%%%%%%%%%%%%%%%%%%%%%%%%%%%%%%%%%%%%%%%%%%%%%%%%%%%%%%%%%%%%%%%%%%%%%%%%%%%%

\section{Equations governing the Alfv\'en wave}\label{appendixB}

Let us  consider the second root to equation (\ref{F1}) which corresponds to the Alfv\'en wave solution, and hence  we assume  $\omega^{2} \neq  \left| {\bf{B}}_0   \right| ^2  \left|   {\bf{k}} \right| ^2$. This simplifies equation (\ref{F1}) to:
\begin{eqnarray}
&&  {\mathcal{F}} \left( \phi, x,y,z,p,q,r \right) =  \omega^2 - \left( {\bf{B}}_0 \cdot {\bf{k}} \right) ^{2} \nonumber\\
 &&  \Rightarrow {\frac{1}{2}} \left[  \omega^2 - \left( B_x p + B_y q  +B_z r \right)^2     \right]=0      \;,     \label{alfvenequation}
\end{eqnarray}
where we have introduced ${1}/{2}$ to simplify the equations later on.

As in $\S\ref{SEC:FAST}$, we can solve this partial differential equation using Charpit's method to reduce the system to seven ordinary differential equations, which can then be solved using, e.g., a Runge-Kutta numerical method. For the Alfv\'en wave solution, Charpit's equations relevant to equation (\ref{alfvenequation}) are:
\begin{eqnarray}
{{\rm{d}}\phi \over {\rm{d}}s}&=&0 \;, \quad {{\rm{d}}t \over {\rm{d}}s}=\omega \;,\quad {{\rm{d}} \omega \over {\rm{d}}s}=0 \;,  \nonumber \\
\frac {dx}{ds} &=& - B_x  { \left( B_x p + B_y q  +B_z r \right)}  \;  ,  \nonumber \\
\frac {dy}{ds} &=& -B_y  { \left( B_x p + B_y q  +B_z r \right)} \;  ,  \nonumber \\
\frac {dz}{ds} &=&  - B_z  { \left( B_x p + B_y q  +B_z r \right)}  \;  ,  \nonumber \\
\frac {dp}{ds} &=& \left( p \frac{\partial B_x}{\partial x} + q \frac{\partial B_y}{\partial x}  + r  \frac{\partial B_z}{\partial x}  \right)\;{ \left( B_x p + B_y q  +B_z r \right)} \;, \nonumber\\
\frac {dq}{ds} &=&    \left( p \frac{\partial B_x}{\partial y} + q \frac{\partial B_y}{\partial y}  + r  \frac{\partial B_z}{\partial y}  \right)\;{ \left( B_x p + B_y q  +B_z r \right)} \;, \nonumber\\
\frac {dr}{ds} &=&   \left( p \frac{\partial B_x}{\partial z} + q \frac{\partial B_y}{\partial z}  + r  \frac{\partial B_z}{\partial z}  \right)\;{ \left( B_x p + B_y q  +B_z r \right)} \;, \label{this_is_the_end}
\end{eqnarray}
where   $p=\frac {\partial \phi} {\partial x}$, $q=\frac {\partial \phi} {\partial y}$, $r=\frac {\partial \phi} {\partial z}$, $B_x$, $B_y$ and $B_z$ are the components of our equilibrium field, $\omega$ is the angular frequency of our wave and  $s$ is the parameter along the characteristic. We note that $\phi={\rm{constant}}=\phi_0$ and $\omega={\rm{constant}}=\omega_0$, i.e. constant angular frequency. Thus,  $t=\omega s+t_0$ and so one can  arbitrarily set $t_0=0$, which corresponds to the leading edge of the wave  starting at $t=0$ when $s=0$.

To generate a planar Alfv\'en  wave launched from $z_0=0.2$, one would then solve equations (\ref{this_is_the_end}) subject to the following initial conditions:
\begin{eqnarray*}
\phi_0&=&0 \;,\quad \omega_0=2\pi \;,\quad -1 \leq x_0 \leq 1 \;,\quad  -1 \leq y_0 \leq 1 \;,\quad z =z_0 \;, \nonumber\\
p_0&=&0 \;,\quad q_0 = 0  \;,\quad r_0 =  \omega_0 / {\left| {B_z\left(x_0,y_0,z_0 \right)} \right|}  \;, \nonumber
\end{eqnarray*}
where we have (arbitrarily) chosen $\omega_0=2\pi$ and $\phi_0=0$. This corresponds to a planar Alfv\'en wave initially at $z=z_0$  and  propagating in the direction of increasing $z$.

%%%%%%%%%%%%%%%%%%%%%%%%%%%%%%%%%%%%%%%%%%%%%%%%%%%%%%%%%%%%%%%%%%%%%%%%%%%%%%%%%%%%%%%%%%%%%%%%%%%%%%%%%%%%%%%%%%%%%%%%%%%%%

\section{${\bf {k}}$ parallel to ${\bf B}_{0}$}\label{appendixA}

In this appendix, we address the scenario $   {\bf {k}} = \lambda       {\bf B}_{0}          $ in which the vectors of our three-dimensional coordinate system   $( {\bf B}_{0},  {\bf k},  {\bf B}_{0} \times {\bf k}$) are no longer linearly independent.  To do this we consider  equation (\ref{F1}):  
\begin{eqnarray}
\frac{\partial^2 } {\partial t^2}{\mathbf{v}}_1 =  \left\{ \nabla \times \left[  \nabla \times \left( {\mathbf{v}}_1 \times {\mathbf{B}}_0 \right) \right] \right\} \times \frac{{\mathbf{B}}_0 }{{{\mu}_0} \rho_0}  \;, \nonumber
\end{eqnarray}
where  we have explicitly included ${{\mu}_0}$ and $\rho_0$. Now assuming  ${\bf {k}} = \lambda       {\bf B}_{0}$ and applying the WKB approximation from equation \ref{WKB}  gives:
\begin{eqnarray*}
\omega ^2 {{{\bf{v}}_1}} &=&    \left\{  {\bf{k}}  \times \left[   {\bf{k}}  \times \left( {{{\bf{v}}_1}}\times {\mathbf{B}}_0 \right) \right] \right\} \times \frac{{\mathbf{B}}_0 }{{{\mu}_0} \rho_0} \\
&=&    \left( {\bf{k}} \cdot {\bf{B}}_0 \right)^2 \frac{{{{\bf{v}}_1}}}{{{\mu}_0} \rho_0}  -   \left( {\bf{k}} \cdot {\bf{B}}_0 \right)   \left( {{{\bf{v}}_1}} \cdot {\bf{B}}_0 \right)\frac{{\bf{k}}}{{{\mu}_0} \rho_0}\\
 &-&  \left( {\bf{k}} \cdot {\bf{B}}_0 \right)   \left( {\bf{k}} \cdot {{{\bf{v}}_1}} \right)\frac{{\bf{B}}_0}{{{\mu}_0} \rho_0} +  \left( {\bf{k}} \cdot {{{\bf{v}}_1}} \right) \left| {\bf{B}}_0 \right|^2 \frac{{\bf{k}}}{{{\mu}_0} \rho_0} \\
&=&   \lambda^2   \frac{\left| {\bf{B}}_0 \right|^2}{{{\mu}_0} \rho_0} \left| {\bf{B}}_0 \right|^2 {{{\bf{v}}_1}}   -  \lambda^2  \frac{\left| {\bf{B}}_0 \right|^2}{{{\mu}_0} \rho_0}   \left( {{{\bf{v}}_1}} \cdot {\bf{B}}_0 \right){\bf{B}}_0 \\
&-&  \lambda^2   \frac{ \left| {\bf{B}}_0 \right|^2}{{{\mu}_0} \rho_0} \left( {\bf{B}}_0 \cdot {{{\bf{v}}_1}} \right){\bf{B}}_0 + \lambda^2  \left( {\bf{B}}_0 \cdot {{{\bf{v}}_1}} \right) \frac{\left| {\bf{B}}_0 \right|^2}{{{\mu}_0} \rho_0} {\bf{B}}_0 \\
&=&    \lambda^2   v_A^2 \left| {\bf{B}}_0 \right|^2 {{{\bf{v}}_1}}   -  \lambda^2  v_A^2  \left( {{{\bf{v}}_1}} \cdot {\bf{B}}_0 \right){\bf{B}}_0 
\end{eqnarray*}
where  $v_A^2={\left| {\bf{B}}_0 \right|^2}/{{{\mu}_0} \rho_0}$. We have explicitly included ${{\mu}_0}$ and $\rho_0$ to make the construction of $v_A^2$ clear.

Thus, for  ${{{\bf{v}}_1}}$ parallel to ${{\bf{{B}}}_0}$, {{i.e.}} ${{{\bf{v}}_1}} =\alpha {{\bf{{B}}}_0}$, we have:
\begin{eqnarray*}
\omega ^2 \alpha {\bf{B}}_0  = \lambda^2   v_A^2 \left| {\bf{B}}_0 \right|^2 \alpha {\bf{B}}_0   -  \lambda^2  v_A^2 \alpha \left|{\bf{B}}_0 \right|^2 {\bf{B}}_0 \quad  \Rightarrow \quad\omega ^2  = 0  \;.
\end{eqnarray*}
So the longitudinal oscillations  (since ${{{{\bf{v}}_1}}}\parallel {{\bf{B}}_0} \parallel {\bf{k}}$) do not propagate, {{i.e.}} this is the dispersion relation for slow waves under the $\beta=0$ assumption.

For ${{{\bf{v}}_1}}$ perpendicular to ${{\bf{B}}_0}$, i.e.  ${{{\bf{v}}_1}} \cdot {{\bf{B}}_0}=0$, which are  transverse oscillations, we have:
\begin{eqnarray*}
\omega ^2 {{{\bf{v}}}}_1 = \lambda^2   v_A^2 \left| {\bf{B}}_0 \right|^2 {{{\bf{v}}}}_1 \quad \Rightarrow \quad \omega ^2 =    v_A^2 \left| {\bf{k}} \right|^2 \;.
\end{eqnarray*}
This is the dispersion relation for a transverse and  incompressible Alfv\'en wave, {{i.e.}}  ${\bf{k}}  \parallel {\bf{B}}_0 \perp {{{\bf{v}}_1}}$, i.e. it is the same as equation (\ref{alfvenequation}) under the assumption ${{\bf{B}}_0} \parallel {\bf{k}}$. However, it is also the dispersion relation for the fast magnetoacoustic wave propagating in the direction of the magnetic field. Thus, we cannot distinguish between these two wave types in this specific scenario.

It is also worth noting that even though the coordinate system we considered in $\S \ref{WKBAPPROXIMATION}$ is not linearly independent when ${{\bf{B}}_0} \parallel {\bf{k}}$, the result from equation (\ref{F1}) still holds. Under the assumption  $   {\bf {k}} = \lambda       {{\bf B}}_{0}          $, equation (\ref{F1}) simplifies to:
\begin{eqnarray*}
\mathcal{F}\left( \phi, x,y,z,p,q,r \right)&=&  \left( \omega  - v_A^2      \left|   {\bf{k}} \right|^2  \right)^2     =0\;.
\end{eqnarray*}
So we have a double root and the solution is degenerate, {{i.e.}} it is impossible to distinguish the waves under these conditions, i.e. this is the same as  equation (\ref{alfvenequation}).

%%%%%%%%%%%%%%%%%%%%%%%%%%%%%%%%%%%%%%%%%%%%%%%%%%%%%%%%%%%%%%%%%%%%%%%%%%%%%%%%%%%%%%%%%%%

\section{Fast wave behaviour in the $y=0$, $xz-$plane and $x=0$, $yz-$plane}\label{appendixC}

We can use the WKB approximation to plot a solution for a wave generated at $-1 \leq x_0 \leq 1 $, $y_0=0$, $z_0=0.2$. This can be seen in  Figure \ref{Figure_appendix_1} which should be compared  to Figure \ref{Figure3} in  $\S\ref{subsec:1}$. We can also use the WKB approximation to plot a solution for a wave generated at $x_0=0$, $-1 \leq y_0 \leq 1 $ and $z_0=0.2$. This can be seen in Figure \ref{Figure_appendix_2} where this should be compared  to Figure \ref{Figure5} in  $\S\ref{subsec:2}$.

%%%%%%%%%%%%%%%%%%%%%%%%%%%%%%%%%%%%%%%%%%%%%%%%%%%%%%%%%%%%%%%%%%%%%%

\begin{figure*}[t]
\begin{center}
\includegraphics[height=16.1cm]{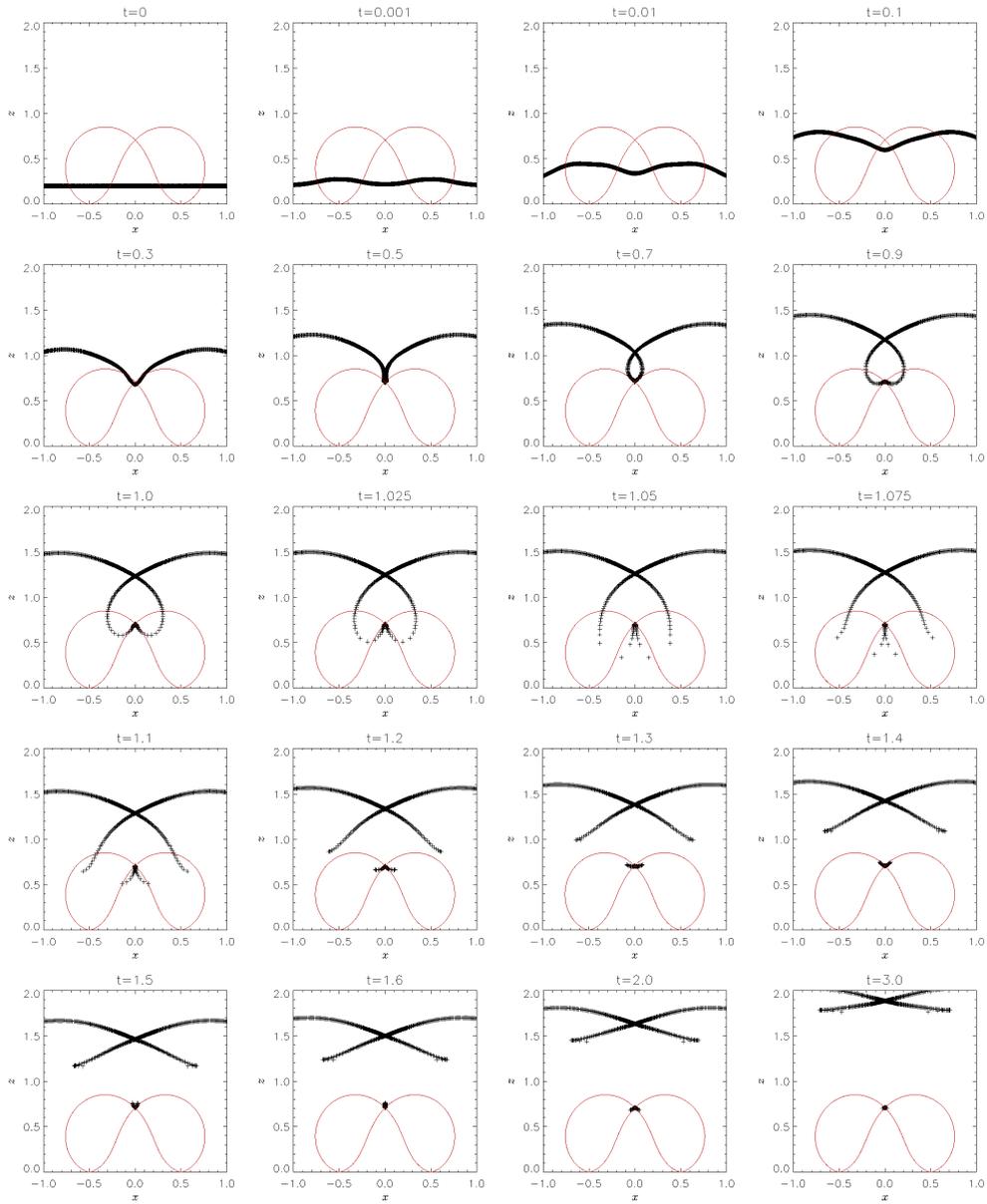}
\caption{Solution of constant values of $\phi$ for WKB approximation of a fast wave generated on lower boundary for $-1 \leq x_0 \leq 1$, $y_0=0$, $z_0=0.2$ and its resultant propagation in the $xz-$plane at various  times.  Displayed times have been chosen to best illustrate evolution so  time between frames is not necessarily uniform. The wavefront  consists of crosses from the WKB wave solution, to better illustrate the evolution. The red separatrices in the $xz-$plane are also shown to provide context.}
\label{Figure_appendix_1}
\end{center}
\end{figure*}

%%%%%%%%%%%%%%%%%%%%%%%%%%%%%%%%%%%%%%%%%%%%%%%

\begin{figure*}[t]
\begin{center}
\includegraphics[height=14cm]{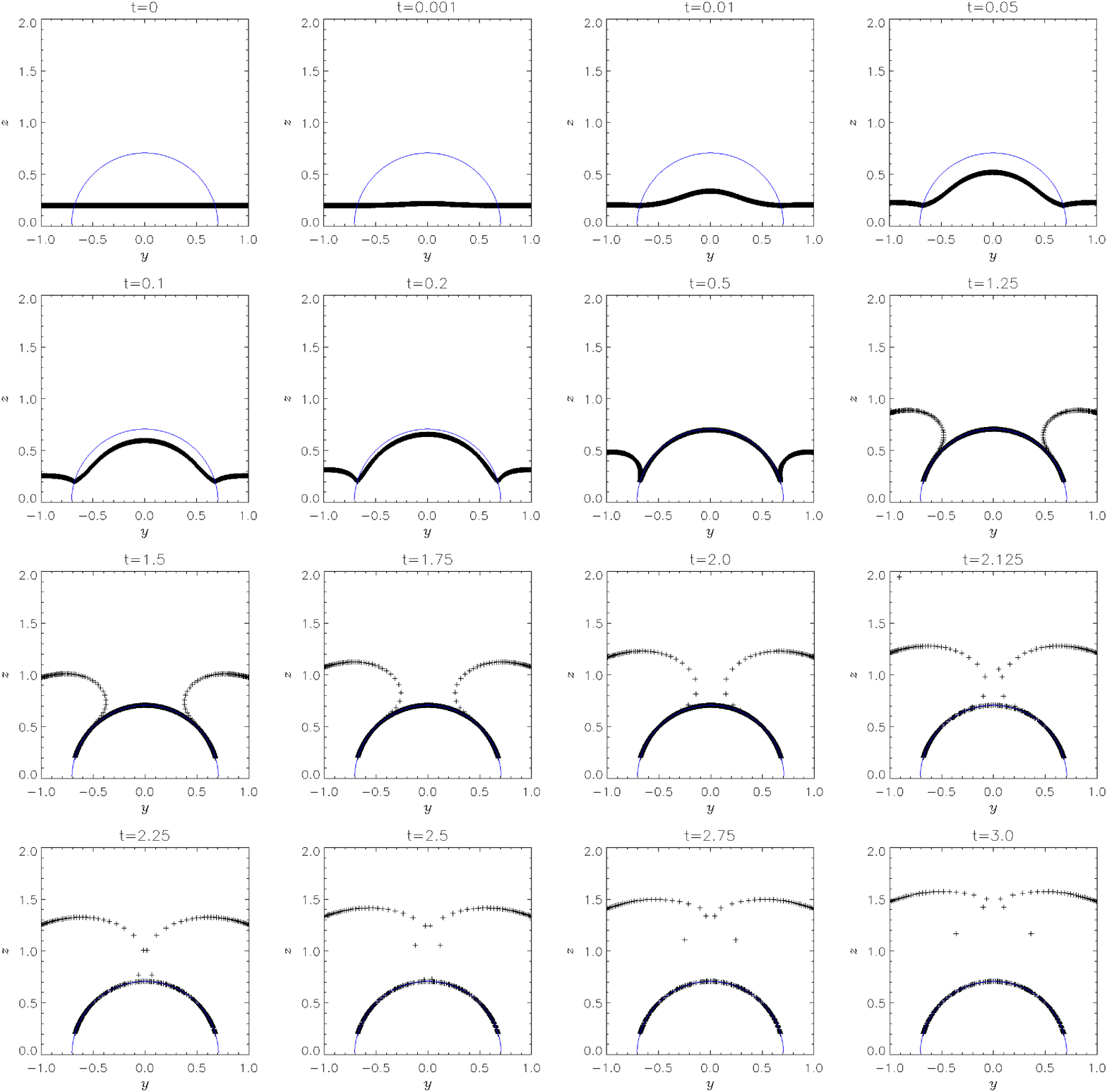}
\caption{Solution of constant values of $\phi$ for WKB approximation of a fast wave generated on lower boundary for $x_0=0$, $-1 \leq y_0 \leq 1$, $z_0=0.2$ and its resultant propagation in the $yz-$plane at  various times.  Displayed times have been chosen to best illustrate evolution so  time between frames is not necessarily uniform. The wavefront  consists of crosses from the WKB wave solution, to better illustrate the evolution.  The blue line indicates the location of the X-line.}
\label{Figure_appendix_2}
\end{center}
\end{figure*}

%%%%%%%%%%%%%%%%%%%%%%%%%%%%%%%%%%%%%%%%%%%%%%%

\end{document}